\documentclass[aps,12pt, a4paper,nofootinbib]{revtex4}

\usepackage[brazil, english]{babel}
\usepackage[utf8]{inputenc}
\usepackage[T1]{fontenc}
\usepackage{amsmath}
\usepackage{amsfonts}
\usepackage{amssymb}
\usepackage{graphics,graphicx}
\usepackage{ulem}
\usepackage{graphicx,color}
\usepackage{subfigure} 
\usepackage{wrapfig} 
\usepackage{epstopdf}
\graphicspath{{figuras/}}

\usepackage{setspace}
\usepackage[unicode=true,bookmarks=false,breaklinks=false,pdfborder={0 0 1},colorlinks=true]
 {hyperref}
\hypersetup{
 citecolor=blue,linkcolor=blue,urlcolor=blue}
\begin{document}

\title{Domain walls in Horndeski gravity}

\author{F. F. Santos}
\email{fabiano.ffs23@gmail.com} \affiliation{Departamento de F\'\i sica, Universidade Federal da Para\'iba, Caixa Postal 5008, 58051-970, Jo\~ao Pessoa, Para\'iba, Brazil.\\
Departamento de Física, Universidade Federal do Maranhão, Campus Universitario do Bacanga, São Luís (MA), 65080-805, Brazil.}

\author{F. A. Brito}
\email{fabrito@df.ufcg.edu.br} \affiliation{Departamento de F\'\i sica, Universidade Federal de Campina Grande, Caixa Postal 10071, 58109-970  Campina Grande, Para\'\i ba,Brazil\\
Departamento de F\'\i sica, Universidade Federal da Para\'iba, Caixa Postal 5008, 58051-970 Jo\~ao Pessoa PB, Brazil}

\begin{abstract}
The Horndeski gravity yields domain wall solutions that connect asymmetric vacua for a certain region of parameters. In the thin wall limit there exist BPS domain walls that connect stable vacua. That is, the false vacua in this theory follows the same effect obtained in supergravity vacua, i.e., it does not decay according to Coleman-de Luccia theory and the BPS solutions interpolating different vacua, e.g., non-degenerate AdS vacua,  are static infinite planar domain walls separating vacua associated with different spacetimes.
\end{abstract}

\maketitle
\newpage

\newpage

\section{Introduction}

According to standard cosmology, the Universe started with an extremely hot and dense state, where all matter and radiation were confined in an infinitely small space. Thus, according to the GUT theories \cite{Nanopoulos:1978rk,Binetruy:1979hc}, as the Universe expands, it cools down and provides the necessary conditions for symmetry breaking, giving the origin to topological defects \cite{Basu:1993rf,Cvetic:1996vr,Dolgov:2016fnx,Dolgov:2017zjp}. Topological defects can occur in a physical system when the vacuum of the system has a non-trivial topology. Furthermore, as discussed by Coleman and De Luccia \cite{Coleman:1980aw} to have a complete description of vacuum decay we cannot neglect the gravitation. This will be well explored in our domain wall solutions with non-degenerate vacua. 

The domain wall is a type of topological defect, which corresponds to the interpolation between two regions separated by different expectation values of a scalar field \cite{Cvetic:1993xe}. Investigations regarding solutions to Einstein's field equations produced by a flat thin domain wall were proposed by Vilenkin \cite{Vilenkin:1981zs}. The idea behind this solution was to perform linear approximation for the gravitational field associated with the aforementioned topological defect. Such a solution has vacuum energy (cosmological constant) that disappears on both sides of the domain wall. This linear solution cannot provide us with any information about the global structure of the gravitational field. It was later shown that the full nonlinear Einstein equations are not satisfied by this solution \cite{Dolgov:1988qw}, but rather satisfied by an exact solution (thin plane domain wall) that has a time-dependent metric that involves a Sitter space along the domain wall geometrical hypersurface \cite{Vilenkin:1984hy}. This should be contrasted with the fact that static flat domain walls can be found as exact solutions of the Einstein equations as long as the cosmological constant of the ambient space is not zero, e.g. static flat domain walls can be achieved as embedded into anti-de-Sitter (AdS) spacetime. {Planar static domain walls that are created in the universe through spontaneously symmetry breaking of a discrete symmetry \cite{Lee} lead to cosmological consequences as noticed long ago \cite{Zeldovich}, since domain walls at the present cosmological horizon would strongly distort the observed isotropy of CMB. In order to solve this problem, several mechanisms addressing destruction of such domain walls were considered in the literature \cite{Kuzmin,Kuzmin1,Kuzmin2,Kuzmin3}.

 Investigations on domain walls have been presented in the context of several modified gravity theories such as Brans–Dicke's theory to probe dilatonic domain walls. This theory provides a static flat domain wall in general relativity coupled to a massless scalar field (see \cite{Groen:1992sm} for further details.) This scalar field that is present in Brans-Dicke's theory develops a solution that plays the role of  the general coupling in theory \cite{La:1992fs}. It is now clear  that the Horndeski gravity accommodates a wide range of dark energy models, such as Lorentz invariant models with a scalar degree of freedom, including for instance, quintessence, $k$-essence, $f(R)$-gravity, Brans-Dicke's theory and Galileons \cite{Charmousis:2011bf,Charmousis:2011ea,Santos:2019ljs,Babichev:2017lmw,Arratia:2020hoy,Starobinsky:2016kua,Bruneton:2012zk,Brito:2018pwe,Brito:2019ose}. These theories have been called attention because the observational cosmological data suggests the need of introducing some mysterious components in the past and current universe to explain its cosmic history in a consistent way \cite{Santos:2019ljs}. The main current interests in these models is the understanding of the accelerated expansion of the Universe. They are essentially grounded on modified theories of gravity. However, there are other candidates to address these issues that are consistent truncations of unification theories such as superstrings/M-theory as for instance supergravity and supersymmetric standard models, which also predict the existence of both high dimensional spacetimes and supersymmetry. Beyond the traditional approach with the extra dimensions extremely small, string theory has suggested a variety of ingredients in extra-dimensional spacetimes that can be useful for the construction of cosmological and phenomenological models \cite{ArkaniHamed:1998rs,Antoniadis:1998ig,Randall:1999ee,Randall:1999vf,Kaloper:1999sm,Nihei:1999mt}. A natural question that is raised is the extent to which there is limit of parameters such that a modified theory of gravity can also enjoy supersymmetric solutions. More specifically, in the following  we shall address the `supergravity type' domain walls in Horndeski gravity.

Thus, we follow recent studies \cite{Brito:2018pwe,Santos:2019ljs}, to apply the first-order formalism to find Bogomol'nyi-Prasad-Sommerfield (BPS) domain wall solutions in Horndeski gravity. Through this formalism, we shall focus on flat domain walls within asymptotically four-dimensional anti-de Sitter (AdS$_{4}$) spacetimes. As in the usual Einstein gravity where a domain wall/cosmology correspondence were stablished \cite{Skenderis:2006jq} --- see also \cite{Bazeia:2007vx} for brane cosmology --- in the sense that the BPS formalism for domain wall solutions can be easily associated with similar formalism for cosmology via analytic continuation, it is interesting to address the same problem in the Horndeski gravity. Thus, we shall consider cosmological fluctuations obtained in Ref.~\cite{Kob} via ADM formalism, to find the gravity fluctuations around domain wall solutions in Horndeski gravity through analytic continuation.

Another interesting point to consider ADM formalism in our analyses is that through this formalism, we show that the Horndeski gravity satisfies $c_{GW}=c_{light}$, which is in agreement with \cite{Bahamonde:2019ipm,Bahamonde:2019shr,Ezquiaga:2017ekz,Creminelli:2017sry,Baker:2017hug}. {In \cite{Charmousis:2011bf,Charmousis:2011ea,Babichev:2017lmw,Arratia:2020hoy} were also considered related models known as self-tuning models, where the energy-momentum tensor of the scalar field almost perfectly balances the large bare cosmological constant, which is assumed to be present in the Lagrangian. Thus, the accelerated observable expansion of the Universe is consistent with a tiny effective cosmological constant.}

The paper is organized as follows. In Sec.~\ref{z1}, we present the Horndeski gravity, and in Sec.~\ref{BBS-section} we develop the first-order formalism in four dimensions and discuss the consequences of non-degenerate AdS vacua. In Sec.~\ref{z2}, we compute domain wall solutions in the context of Horndeski gravity through numerical methods. In Sec.~\ref{z3}, we consider cosmological tensor perturbations through the ADM formalism \cite{Kob}  to  find its analytical continued counterpart for domain walls. Finally, in Sec.~\ref{z4}, we present our final comments.

\section{Horndeski gravity}\label{z1}

In this section, we investigate the presence of domain walls in Horndeski gravity. For this, we perform the equations of motion with a domain walls {\it Ansatz} as pointed in \cite{Bazeia:2007vx}. In fact, there are many scenarios presenting domain walls solutions in Einstein gravity in the literature \cite{Cvetic:1996vr,Vilenkin:1981zs,Vilenkin:1984hy,Lee,Zeldovich,Kuzmin,Kuzmin1,Kuzmin2,Kuzmin3,Groen:1992sm,La:1992fs,Bazeia:2007vx}. But, here in our case, this is the first time that we talk about this procedure in the theory beyond Einstein gravity. Horndeski in \cite{Horndeski:1974wa} proposes a theory with the following Lagrangian
\begin{eqnarray}
&&\mathcal{L}_{H}=\mathcal{L}_{2}+\mathcal{L}_{3}+\mathcal{L}_{4}+\mathcal{L}_{5},\label{01}\\
&&\mathcal{L}_{2}=G_{2}(X,\phi),\nonumber\\
&&\mathcal{L}_{3}=-G_{3}(X,\phi)\Box\phi,\nonumber\\
&&\mathcal{L}_{4}=G_{4}(X,\phi)R+\partial_{X}G_{4}(X,\phi)\delta^{\mu\nu}_{\alpha\beta}\nabla^{\alpha}_{\mu}\phi\nabla^{\beta}_{\nu}\phi,\nonumber\\
&&\mathcal{L}_{5}=G_{5}(X,\phi)G_{\mu\nu}\nabla^{\mu\nu}\phi-\frac{1}{6}\partial_{X}G_{5}(X,\phi)\delta^{\mu\nu\rho}_{\alpha\beta\gamma}\nabla^{\alpha}_{\mu}\phi\nabla^{\beta}_{\nu}\phi\nabla^{\gamma}_{\rho}\phi,\nonumber
\end{eqnarray}
where ${X\equiv -\frac{1}{2}(\nabla\phi)^{2}}$ is the standard kinematic term. The Lagrangian of this model with the single scalar field provides second-order field equations and second-order energy-momentum tensor \cite{Horndeski:1974wa,Anabalon:2013oea,Cisterna:2014nua,Heisenberg:2018vsk}. In Eq.~\eqref{01}, we can note there are arbitrary $G_{k}(X,\phi)$ functions of the scalar field and its kinetic term. {However, Horndeski's theory is difficult to analyze without specifying these functions. There is a special subclass of this theory, sometimes called the Fab Four (F4) \cite{Charmousis:2011bf,Charmousis:2011ea}, for which the coefficients are chosen such that the Lagrangian is written in the form
\begin{eqnarray}
&&\mathcal{L}_{F4}=\mathcal{L}_{John}+\mathcal{L}_{Paul}+\mathcal{L}_{George}+\mathcal{L}_{Ringo }-2\Lambda,\\
&&\mathcal{L}_{John}=U_{John}(\phi)G_{\mu\nu}\nabla^{\mu}\phi\nabla^{\nu}\phi,\\
&&\mathcal{L}_{Paul}=U_{Paul}(\phi)P_{\mu\nu\rho\sigma}\nabla^{\mu}\phi\nabla^{\rho}\phi\nabla ^{\nu\sigma}\phi,\\
&&\mathcal{L}_{George}=U_{George}(\phi)R,\\
&&\mathcal{L}_{Ringo}=U_{Ringo}(\phi)(R_{\mu\nu\alpha\beta}R^{\mu\nu\alpha\beta}-4R_{\mu\nu }R^{\mu\nu}+R^{2}).
\end{eqnarray}
From these equations, we can see that any self-tuning scalar-tensor theory, which is based on the equivalence principle, must be built from the Fab Four (F4). However, the weakest of the four terms is Ringo, due to the fact that it cannot bring about self-tuning without ``a little help from his friends'', John and Paul. So in this case, as discussed by \cite{Charmousis:2011bf,Charmousis:2011ea} the Ringo has a non-trivial effect on cosmological dynamics, but it does not spoil the self-tuning. George also has difficulties to go it alone, in that sense when $U_{George}=$const., we only have General Relativity and we do not have self-tuning. However, when $U_{George}\neq$const., we have Brans-Dicke gravity, which develops self-tuning. As such, it is expected that one should always include John and/or Paul for the reasons given, and because their non-trivial derived interactions can give rise to Vainshtein effects that would help pass the solar system tests \cite{Charmousis:2011bf}. Here the double dual of the Riemann tensor is given by
\begin{eqnarray}
P^{\mu\nu}_{\alpha\beta}=-\frac{1}{4}\delta^{\mu\nu\gamma\delta}_{\sigma\lambda\alpha\beta}R ^{\sigma\lambda}_{\gamma\delta}=-R^{\mu\nu}_{\alpha\beta}+2R^{\mu}_{[\alpha}\delta^{\nu }_{\beta ]}-2R^{\nu}_{[\alpha}\delta^{\mu}_{\beta ]}-R\delta^{\mu}_{[\alpha}\delta ^{\nu}_{\beta ]}.
\end{eqnarray}
The contraction $P^{\mu\nu}_{\nu\alpha}=G^{\mu}_{\alpha}$ is the Einstein tensor. A particular model related to the F4 theories can be obtained by setting the potential functions to constant values, $U_{J}=\eta/2$, $U_{P}=U_{R}=0$, $U_{G} =\kappa$ and adding an extra term, the standard kinetic term X to the scalar field, which gives the model that corresponds to the John's Lagrangian \cite{Starobinsky:2016kua} and finally by adding a scalar potential $\tilde{V}(\phi)$ \cite{Charmousis:2011bf,Charmousis:2011ea,Babichev:2017lmw,Starobinsky:2016kua,Bruneton:2012zk}, we have {the Horndeski-like gravity} action of interest in the form}
\begin{equation}
S[g_{\mu\nu},\phi]=\int{\sqrt{-g}d^{4}x\left[\kappa R-\frac{1}{2}(\alpha g_{\mu\nu}-\eta G_{\mu\nu})\nabla^{\mu}\phi\nabla^{\nu}\phi-\tilde{V}(\phi)\right]}.\label{1}
\end{equation}
Here $\kappa=(16\pi G)^{-1}$. {The scalar field has dimension of $mass$}, and the parameters $\alpha$ and $\eta$ control the strength of the kinetic couplings, $\alpha$ is dimensionless and $\eta$ has dimension of $(mass)^{-2}$. We now compute the Einstein-Horndeski field equations that read\footnote{We shall adopt $4\pi G=1$ such that $\kappa=1/4$ along the text of the paper.} 
\begin{equation}
G_{\mu\nu}=\frac{1}{2\kappa}T_{\mu\nu},\label{2}
\end{equation}
with 
\begin{equation}\label{Tmn}
T_{\mu\nu}=\alpha T^{(1)}_{\mu\nu}-g_{\mu\nu}\tilde{V}(\phi)+\eta T^{(2)}_{\mu\nu},
\end{equation}
and the scalar field equation is given by
\begin{equation}
\nabla_{\mu}J^{\mu}=\tilde{V}_{\phi};\quad J^{\mu}=(\alpha g^{\mu\nu}-\eta G^{\mu\nu})\nabla_{\nu}\phi.\label{3}
\end{equation}
The energy-momentum tensors $T^{(1)}_{\mu\nu}$, $T^{(2)}_{\mu\nu}$ take the following form
\begin{eqnarray}
T^{(1)}_{\mu\nu}&=&\nabla_{\mu}\phi\nabla_{\nu}\phi-\frac{1}{2}g_{\mu\nu}\nabla_{\lambda}\phi\nabla^{\lambda}\phi,\label{4}\\
T^{(2)}_{\mu\nu}&=&\frac{1}{2}\nabla_{\mu}\phi\nabla_{\nu}\phi R-2\nabla_{\lambda}\phi\nabla_{(\mu}\phi R^{\lambda}_{\nu)}-\nabla^{\lambda}\phi\nabla^{\rho}\phi R_{\mu\lambda\nu\rho}\nonumber\\
              &-&(\nabla_{\mu}\nabla^{\lambda}\phi)(\nabla_{\nu}\nabla_{\lambda}\phi)+(\nabla_{\mu}\nabla_{\nu}\phi)\Box\phi+\frac{1}{2}G_{\mu\nu}(\nabla\phi)^{2}\nonumber\\
							&-& g_{\mu\nu}\left[-\frac{1}{2}(\nabla^{\lambda}\nabla^{\rho}\phi)(\nabla_{\lambda}\nabla_{\rho}\phi)+\frac{1}{2}(\Box\phi)^{2}-(\nabla_{\lambda}\phi\nabla_{\rho}\phi)R^{\lambda\rho}\right].\label{5}							
\end{eqnarray}
The metric that describes solutions of two-dimensional flat domain walls within asymptotically four-dimensional Minkowski (M$_{4}$) or Anti-de Sitter (AdS$_{4}$) spacetimes  can be found by using the {\it Ansatz} \cite{Cvetic:1996vr}
\begin{equation}
ds^{2}=dy^{2}+a^{2}(y)(-dt^{2}+dx^{2}+dz^{2}).\label{6}
\end{equation}
Below we write two equations from the Einstein-Horndeski field equations \eqref{2} that in terms of the metric Ansatz are given by
\begin{equation}
\tilde{H}^{2}(y)=\frac{\alpha\phi'^{2}(y)-2\tilde{V}(\phi)}{3(4\kappa+3\eta\phi'^{2}(y))},\label{7}
\end{equation}
and
\begin{eqnarray}
&&{\frac{a^{'2}(y)}{a^{2}(y)}\left(4\kappa+\eta\psi^{2}(y)\right)+2\frac{a^{''}(y)}{a(y)}(4\kappa+\eta\psi^{2}(y))+}\nonumber\\
&&{4\eta\psi(y)\frac{a^{'}(y)}{a(y)}\psi^{'}(y)+2\tilde{V}(\psi(y))+\alpha\psi^{2}(y)=0},\label{7-8}
\end{eqnarray}
where we have introduced a new field
\begin{equation}
{\phi^{'}(y)=\frac{d\phi(y)}{dy}\equiv\psi(y)}.
\end{equation}
The Klein-Gordon equation can be obtained through the scalar field equation (\ref{3}), which is written as
\begin{equation}
\phi^{''}+3\tilde{H}\phi^{'}-\frac{6\eta\phi^{'}\tilde{H}\tilde{H}^{'}}{\alpha-3\eta \tilde{H}^{2}}-\frac{\tilde{V}_{\phi}(\phi)}{\alpha-3\eta\tilde{H}^{2}}=0.\label{9}
\end{equation}

In order to address the issues considered in the next section we shall now introduce the first order formalism presented as follows\footnote{For more details about this procedure in Horndeski gravity see \cite{Brito:2018pwe,Santos:2019ljs}.}
\begin{eqnarray}
\tilde{H}\equiv a{'}/a\longrightarrow &&A'=-\tilde{W}(\phi)/(4\kappa),\cr
&&\phi{'}=+\tilde{W}_{\phi}(\phi),\label{fist}
\end{eqnarray}
where for convenience we assume $a(y)=\exp A(y)$. Through these equations, the scalar potential is consistent with the Einstein-Horndeski equations if written as\footnote{Note that the scalar potential in terms of the superpotential is similar to the cases \cite{Brito:2018pwe,Santos:2019ljs}.}
\begin{equation}
\tilde{V}(\phi)=\frac{\alpha}{2}\tilde{W}_{\phi}^{2}-\frac{3}{2}\left(\frac{\tilde{W}}{4\kappa}\right)^{2}(4\kappa+3\eta\tilde{W}^{2}_{\phi}).\label{8}
\end{equation}
Now combining the Eq.~(\ref{8}) with Eq.~(\ref{9}), we have an important constraint on the `superpotential' 
\begin{eqnarray}
2\tilde{W}\tilde{W}_{\phi\phi}+\tilde{W}_{\phi}^{2}+3\frac{\tilde{W}^{2}}{4\kappa}-\beta=0,\label{10} 
\end{eqnarray}
where $\beta=\kappa(\alpha-1)/\eta$. In the cosmological scenarios obtained via analytical continued domain wall solutions one should replace $\beta\to -\beta$, that means changing $\eta\to -\eta$. In the next section, we shall solve the above equation by numerical methods.

\section{The BPS solutions}
\label{BBS-section}

We are considering an extended theory of gravity. General relativity theory with suitable couplings to matter fields can produce supergravity as well-known since 1970's. However, this is not clear in general for extended gravity theories. Then, one should investigate at which limits the extended theory of gravity or even some models of usual gravity theory can reach supergravity limits. In general grounds, these are considered ``fake supergravity'' theories. In our set-up this is not different. We should explain at which regimes we are indeed close to a `fake supergravity' theory.

The equations of motion given by Eqs.~\eqref{7}, \eqref{7-8}, and \eqref{9} are not independent since we can combine two of them to find the third one. Thus, let us focus on Eqs.~\eqref{7-8} and \eqref{9}. They can be obtained from the following energy-functional
\begin{eqnarray}
E[A,\phi]=\int{e^{3A}dy\left[-\frac{1}{2}\hat{\phi}'^2+\frac32 A'^2+\frac32\hat{\eta}\hat{\phi}'^2A'^2-\tilde{V}(\hat{\phi})\right]},\label{EF}
\end{eqnarray}
where, by convenience, we have also made the following replacements $\phi=\hat{\phi}/\sqrt{\alpha}$ and $\eta=\hat{\eta}\alpha$. Now by considering the potential defined in \eqref{8} in terms of hatted quantities we can rewrite \eqref{EF} as follows
\begin{eqnarray}
E[A,\hat{\phi}]&=&\int{e^{3A}dy\left[-\frac{1}{2}\hat{\phi}'^2+\frac32 A'^2+\frac32\hat{\eta}\hat{\phi}'^2A'^2-\frac{1}{2}\tilde{W}_{\hat\phi}^{2}+\frac{3}{2}\tilde{W}^{2}(1+3\hat{\eta}\tilde{W}^{2}_{\hat\phi})\right]},\nonumber\\
&=&\int{e^{3A}dy\left\{-\frac{1}{2}\left[(\hat{\phi}'-\tilde{W}_{\hat\phi})^2+2\hat{\phi}'\tilde{W}_{\hat\phi}\right]+\frac32 \left[(A'+\tilde{W})^2-2A'\tilde{W}\right]\right.}\nonumber\\
&+&\left.\frac32\hat{\eta}\hat{\phi}'^2A'^2+\frac{9}{2}\hat{\eta}\tilde{W}^{2}\tilde{W}^{2}_{\hat\phi}\right\}.\label{EF2} 
\end{eqnarray} 
Under the first order equations \eqref{fist} by using the identity $A''=-\tilde{W}'(\hat{\phi})\to A''=-\tilde{W}_{\hat\phi}\hat{\phi}' =-\tilde{W}_{\phi}{\phi}' = - \tilde{W}_{\phi}^2=-{\phi}'^2$ we find the extremized functional given by 
\begin{eqnarray}
E[A,\phi]\!=\!\int{e^{3A}dy\left[A''+3A'^2+6\hat{\eta}\tilde{W}^{2}\tilde{W}^{2}_{\hat\phi})\right]}\!=\!-e^{3A}\tilde{W}({\phi})\Big{|}_{bdry}-6\hat{\eta}\int{e^{3A}dyA'^{2}A''}.\label{EF3}
\end{eqnarray}
Since the above expressions descend from \eqref{EF} whose integrand of it is precisely the effective Lagrangian where the domain wall Ansatz was introduced, we conclude that Eqs.~\eqref{EF2}-\eqref{EF3} express nothing but the domain wall surface density $\sigma$ with opposite sign. Thus, we find the famous Bogomol'nyi bound for the surface density
\begin{eqnarray}
\sigma\geq
e^{3A}\tilde{W}({\phi})\Big{|}_{bdry}+6\hat{\eta}\int{e^{3A}dyA'^{2}A''}.\label{EF3-sigma}
\end{eqnarray}

\subsection{Euclidean action}

Let us first consider the Lagrangian for scalar fields depending only on one spatial coordinate, say $y$ as suggested by the metric \eqref{6}. So the scalar fields and the geometrical functions can be written explicitly in the action \eqref{1} as follows
\begin{equation}
S[A,\phi]=\int{d^{4}x\,e^{3A}\left[ -\left(3A'^2+\frac32A''\right)-\frac{1}{2}\alpha\phi'^2+\frac12\eta\phi'^2(-3A'^2)-\tilde{V}(\phi)\right]},\label{1-2}
\end{equation}
where prime means derivative with respect to the coordinate  $y$. 

Since the theory we are considering here presents large possibilities of  BPS solutions we are going to investigate a more general type of solution other than simple flat domain wall solutions. As we shall see later explicitly, the `supergravity vacua' are not always degenerated so bubble walls related to vacuum decay may appear. Thus, as usual in such  systems, it is more convenient to deal with the regularized Euclidean action, so we rewrite \eqref{1-2} as follows
\begin{equation}
S_E[A,\phi]=\int{d^{4}x\,e^{3A}\left[ 3A'^2+\frac32A''+\frac{1}{2}\alpha\phi'^2+\frac12\eta\phi'^2(-3A'^2)+\tilde{V}(\phi)\right]}+S_{bdry},\label{SE}
\end{equation}
where $S_{bdry}$ is the Euclidean Gibbons-Hawking boundary action \cite{Masoumi:2016pqb}. After algebraic manipulations we recast the Euclidean action as
\begin{eqnarray}
S_E[A,\phi]&=&
\int d^{4}x\, e^{3A}\left[ \frac32A'^2+\frac32(A'^2+A'')+\frac{1}{2}\alpha\phi'^2
-\frac32\eta\phi'^2A'^2+\tilde{V}(\phi)\right] +S_{bdry},\nonumber\\
&=&\int d^{4}x\, e^{3A}\left[ \frac{1}{2}\alpha\phi'^2 -\frac32A'^2
-\frac32\eta\phi'^2A'^2+\tilde{V}(\phi)\right],\label{SE2} 
\end{eqnarray}
where a surface term that comes from the integration by parts  is exactly canceled by the Euclidean Gibbons-Hawking term $S_{bdry}=-(3/2)\exp{(3A)}\,A'\,|_{bdry}$. 

Notice that the integrand in \eqref{SE2} is precisely the same as in equation \eqref{EF} with opposite sign, up to redefinition of $\phi$ and $\eta$. Thus, we can take advantage of Eq.~\eqref{EF3} to rewrite the Euclidean action as
\begin{eqnarray}\label{SE-3}
S_E[A,\phi]&=&\int d^{3}x\Big(e^{3A}\tilde{W}({\phi})\Big{|}_{y_1}^{y_2}+6\hat{\eta}\int_{y_1}^{y_2}{e^{3A}dyA'^{2}A''} \Big).
\end{eqnarray}
This action enables us to compute the probability of bubble nucleation via decay of the false vacuum $\phi(y_2)\equiv\phi^{(2)}_{\rm vac}$ into the true vacuum $\phi(y_1)\equiv\phi^{(1)}_{\rm vac}$ across the nucleation rate per unit volume given by $\Gamma\sim e^{-B}$, where $B=S_E(\phi_{\rm bounce})-S_E(\phi^{(2)}_{\rm vac})$ and the bounce is the $O(4)$ symmetric solution of Euclidean equations of motion that interpolates between the true and false vacuum. In the thin wall limit, the wall of the bounce solution, which is a spherical domain wall, tends to an infinite planar domain wall. 

\subsection{Thin wall limit}\label{thin-wall-limit-subs}

We can under certain conditions assume that the bubble wall is very thin in comparison with the bubble radius. In the case of both false and true vacua are AdS vacua, then we compare the bubble wall thickness with the AdS radii. Thus, let us first recast the last term of Eq.~\eqref{SE-3} by means of the `bubble radius' $\rho(y)=\exp {A(y)}$ such that we find
\begin{eqnarray}\label{last-term-SE-3}
2\int_{y_1}^{y_2}{e^{3A}(A'^{3})'dy}=2\int_{y_1}^{y_2}{\rho^3\left[\frac{3\rho'^2\rho''}{\rho^3}-3\frac{\rho'^3}{\rho^4} \right]dy}\simeq-6\int_{y_1}^{y_2}{e^{2A}A'^{3}dy},
\end{eqnarray}
where we have assumed that $\rho$ evolves linearly inside the interval of integration. This is in accord with the fact that for a bubble wall sufficiently thin, the warp function $A$ is approximately linear in this region. So, by considering that $A(y)=k_i y +{\cal O}(y^2)$, for $k_i=k_1$ in the interval $y_1\leq y\leq y_0$ and $k_i=k_2$ in the interval $y_0\leq y\leq y_2$ we find
\begin{eqnarray}\label{last-term-2-SE-3}
\int_{y_1}^{y_2}{e^{2A}A'^{3}dy}&\simeq& k_1^{3}\int_{y_1}^{y_0}{e^{2k_1y}\, dy}+k_2^{3}\int_{y_0}^{y_2}{e^{2k_2y}\, dy}\nonumber\\
&=&\frac12 k_1^2(e^{2k_1y_0}-e^{2k_1y_1})+\frac12 k_2^2(e^{2k_2y_2}-e^{2k_2y_0})\nonumber\\
&\simeq&(k_1^3+k_2^3)\frac{\Delta}{2}=\left(\frac{1}{\ell_{AdS_{(1)}}^3}+\frac{1}{\ell_{AdS_{(2)}}^3}\right)\frac{\Delta}{2}.
\end{eqnarray}
In the intermediate step we have assumed a linear approximation in the exponentials, whereas in the last step we have identified $y_0-y_1\approx \Delta/2$, $y_2-y_0\approx\Delta/2$ and $k_i\equiv W(\phi^{i}_{\rm vac})=1/\ell_{AdS_{(i)}}$, that are AdS radii. Finally, we find that the thin wall limit is achieved as $\Delta(1/\ell_{AdS_{(1)}}^3+1/\ell_{AdS_{(2)}}^3)\ll1$. At this regime the last
term of Eq.~\eqref{SE-3} is very small compared with the first term and can be neglected.

At thin wall limit the contributions of the tunneling exponent can be split into three different contributions  $B=B_{\rm interior}+B_{\rm wall}+B_{\rm exterior}$ with respect to the bubble. To compute $B_{\rm wall}$ we can now use Eq.~\eqref{SE-3} restricted only to the first term to write 
\begin{eqnarray}\label{SE-4}
B_{\rm wall}=S_E(\phi_{\rm bounce})-S_E(\phi^{(2)}_{\rm vac})=\int d^{3}x\Big(e^{3A}\tilde{W}({\phi})\Big{|}_{y_1}^{y_2}-e^{3A}\tilde{W}(\phi^{(2)}_{\rm vac})\Big{|}_{y_1}^{y_2}\Big).
\end{eqnarray}
As in the previous discussions with respect to the last term of  Eq.~\eqref{SE-3}, it is interesting to consider the radius $\rho(y)=\exp A(y)$ to make further considerations on the above expression that can be rewritten as
\begin{eqnarray}\label{SE-5}
B_{\rm wall}&=&\int d^{3}x\Big(\rho^{3}\tilde{W}({\phi})\Big{|}_{y_1}^{y_2}-\rho^{3}\tilde{W}(\phi^{(2)}_{\rm vac})\Big{|}_{y_1}^{y_2}\Big)\nonumber\\
&=&\int d^{3}x\Big(\rho_2^{3}\tilde{W}(\phi^{(2)}_{\rm vac})-\rho_1^{3}\tilde{W}(\phi^{(1)}_{\rm vac})-(\rho_2^{3}-\rho_1^{3})\tilde{W}(\phi^{(2)}_{\rm vac})\Big).
\end{eqnarray}
Notice that the bounce term can be expressed in the form
\begin{eqnarray}\label{SE-6}
\rho_2^{3}\tilde{W}(\phi^{(2)}_{\rm vac})-\rho_1^{3}\tilde{W}(\phi^{(1)}_{\rm vac})&=&(\rho_2^3-\rho_1^3)\tilde{W}(\phi^{(2)}_{\rm vac})+(\tilde{W}(\phi^{(2)}_{\rm vac})-\tilde{W}(\phi^{(1)}_{\rm vac}))\rho_1^3,\nonumber\\
&=&(\rho_2^3-\rho_1^3)\tilde{W}(\phi^{(2)}_{\rm vac})+\rho_1^3\int_{y_1}^{y_2}{dy\frac{d\tilde{W}}{dy}},\nonumber\\
&=&(\rho_2^3-\rho_1^3)\tilde{W}(\phi^{(2)}_{\rm vac})+\rho_1^3\int_{y_1}^{y_2}{\!\!\!dy\, \phi'^2}.
\end{eqnarray}
In the last step we considered the Eqs.~\eqref{fist} to rewrite the derivative of the superpotential in terms of the $\phi'^2$. Now substituting \eqref{SE-6} into \eqref{SE-5} we find \cite{Masoumi:2017vip}
\begin{eqnarray}\label{SE-7}
B_{\rm wall}=2\pi^2\rho_1^3\int_{y_1}^{y_2}{\!\!\!dy\, \phi'^2},
\end{eqnarray}
where we have written the volume element  of the 4-dimensional Euclidean action in terms of the surface area of a  3-sphere of radius $\rho_1$. Thus, we conclude that the bounce wall contribution is related to the surface tension 
\begin{eqnarray}\label{SE-8}
\sigma&=&\int_{y_1}^{y_2}{\!\!\!dy\, \phi'^2},\nonumber\\
&=&\tilde{W}_2-\tilde{W}_1=\sigma_c,
\end{eqnarray}
where $\tilde{W}_i=\tilde{W}(\phi^{(i)}_{\rm vac})$ and $\sigma_c$ is the critical surface tension. Notice this violates the Coleman-de Luccia (CdL) \cite{Coleman:1980aw}  bound $\sigma<\sigma_c$. We shall discuss further consequences about this shortly. This is certainly not the case in `thick wall' regime, where this condition of violability of the CdL bound does not happen because the $\eta$-term in Eq.~\eqref{SE-3} will contribute and a bounce solution from the full Euclidean action can appear.

\section{Numerical results}
\label{z2}

In this section, we present our numerical results as the solutions of the pair of first order equations in \eqref{fist}. {Such quantities are solved by numerical methods considering a range of values as long as we assume appropriate boundary conditions:
Firstly, we solve \eqref{10} with $\tilde{W}(0)=1$, $\tilde{W}'(0)=1$ via a numerical code that generates outputs used in the next stage of the numerical integration. These boundary conditions is due to \eqref{10} that can easily satisfy them for arbitrary values of $\beta$. Secondly, under the use of the previous outputs we solve \eqref{fist}. To find domain walls solutions, at $y=0$, it is natural that $\phi(0)=0$ and $\phi' (0)=1$. In this case we use the Euler method to  numerically   integrate the first-order differential equations \eqref{fist}.} With these boundary conditions we draw the domain wall solutions and their respective scalar potentials with the vacua structure connected by these domain walls ---  
see Fig.~\ref{p0-1} (for $\beta=2.273$, with $\alpha=5/2$, $\eta=0.66$ (blue) and $\beta=3.125$, with $\alpha=5/2$, $\eta=0.48$ (red)). The geometric behavior formed due to the domain walls are depicted in Fig.~\ref{p0-2} both in terms of the warp factor and the scale factor. The geometry precisely follows a type III domain walls investigated in \cite{Cvetic:1996vr}. Another fact that lead us to identify the type of these domain walls is that $\tilde{W}(\phi)$ never cross the origin --- see Fig.~\ref{W-Wp}.  It clear connects AdS-AdS vacua as can be seen from the vacua at the scalar potentials in Fig.~\ref{p0-1} --- the local minima on the left side to local maxima on the right side.

\begin{figure}[!ht]
\begin{center}
\includegraphics[scale=0.35]{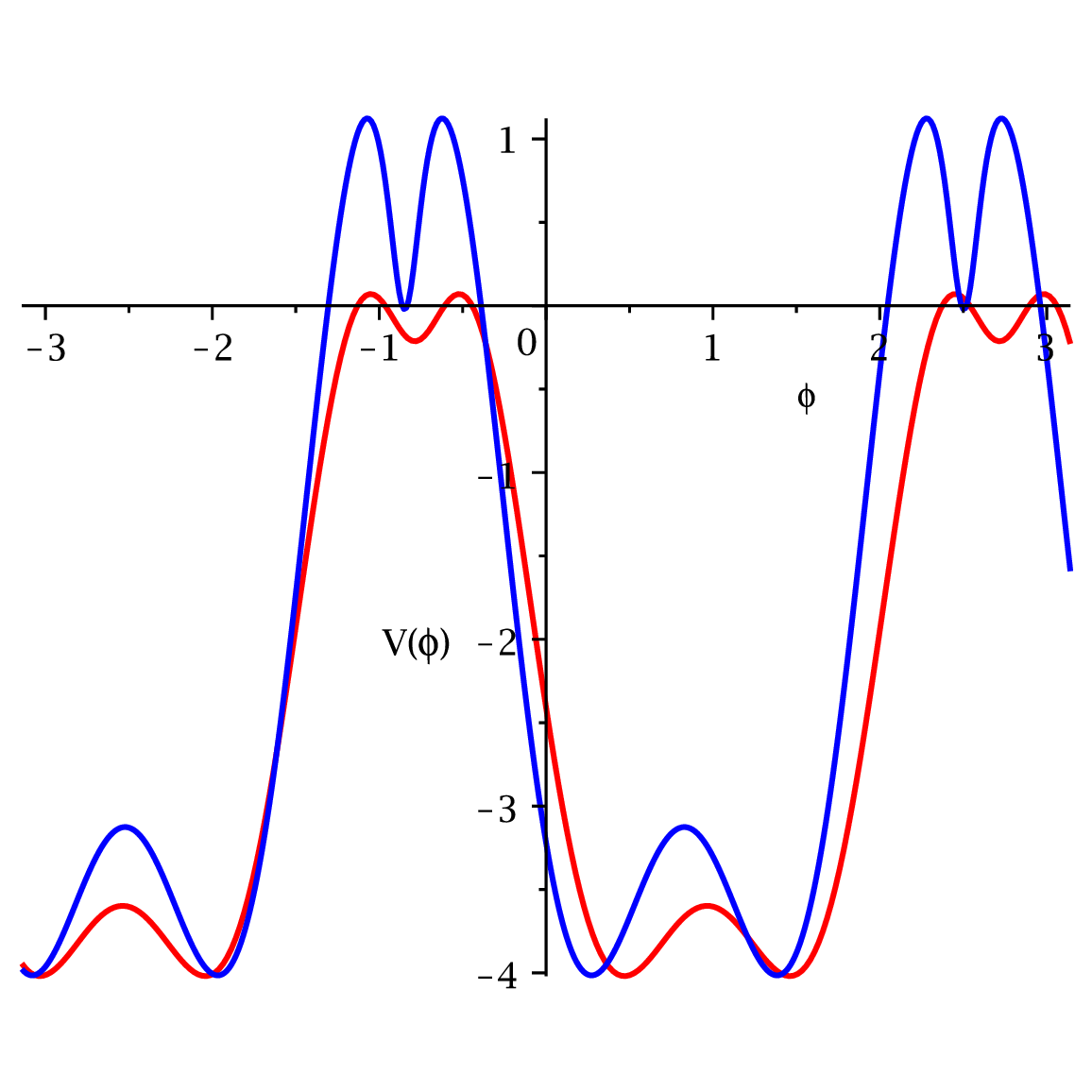}\qquad\qquad\qquad
\includegraphics[scale=0.35]{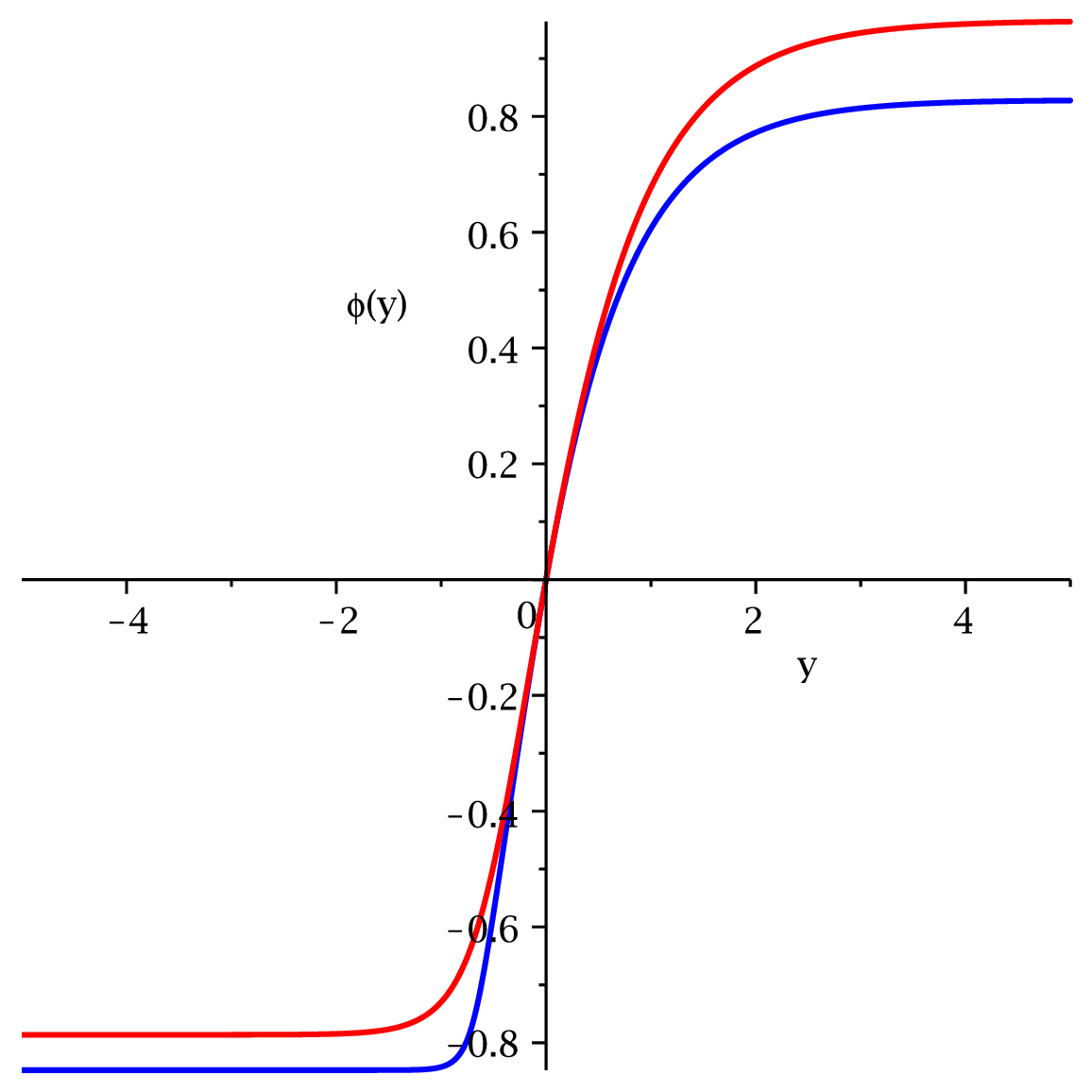}
\caption{The figure shows the scalar potential (\ref{8}) and kinks connecting the false vacua for the values $\beta=2.273$, with $\alpha=5/2$, $\eta=0.66$ (blue) and $\beta=3.125$, with $\alpha=5/2$, $\eta=0.48$ (red). For explicit numerical values of the vacua sector $\phi_\pm \in (-1,1)$, see Table \ref{Tb1}.}
\label{p0-1}
\end{center}
\end{figure}

\begin{figure}[!ht]
\begin{center}
\includegraphics[scale=0.35]{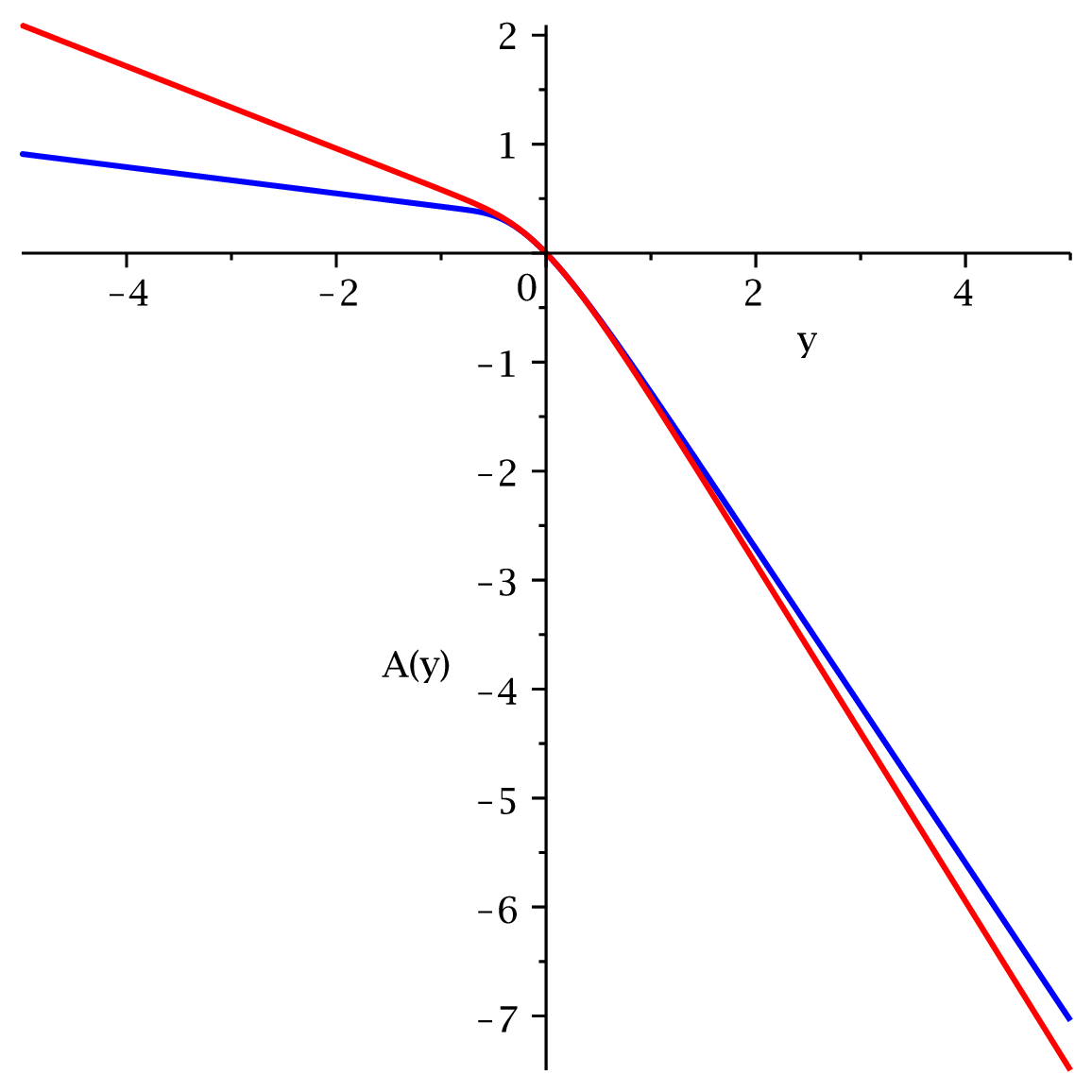}\qquad\qquad\qquad
\includegraphics[scale=0.35]{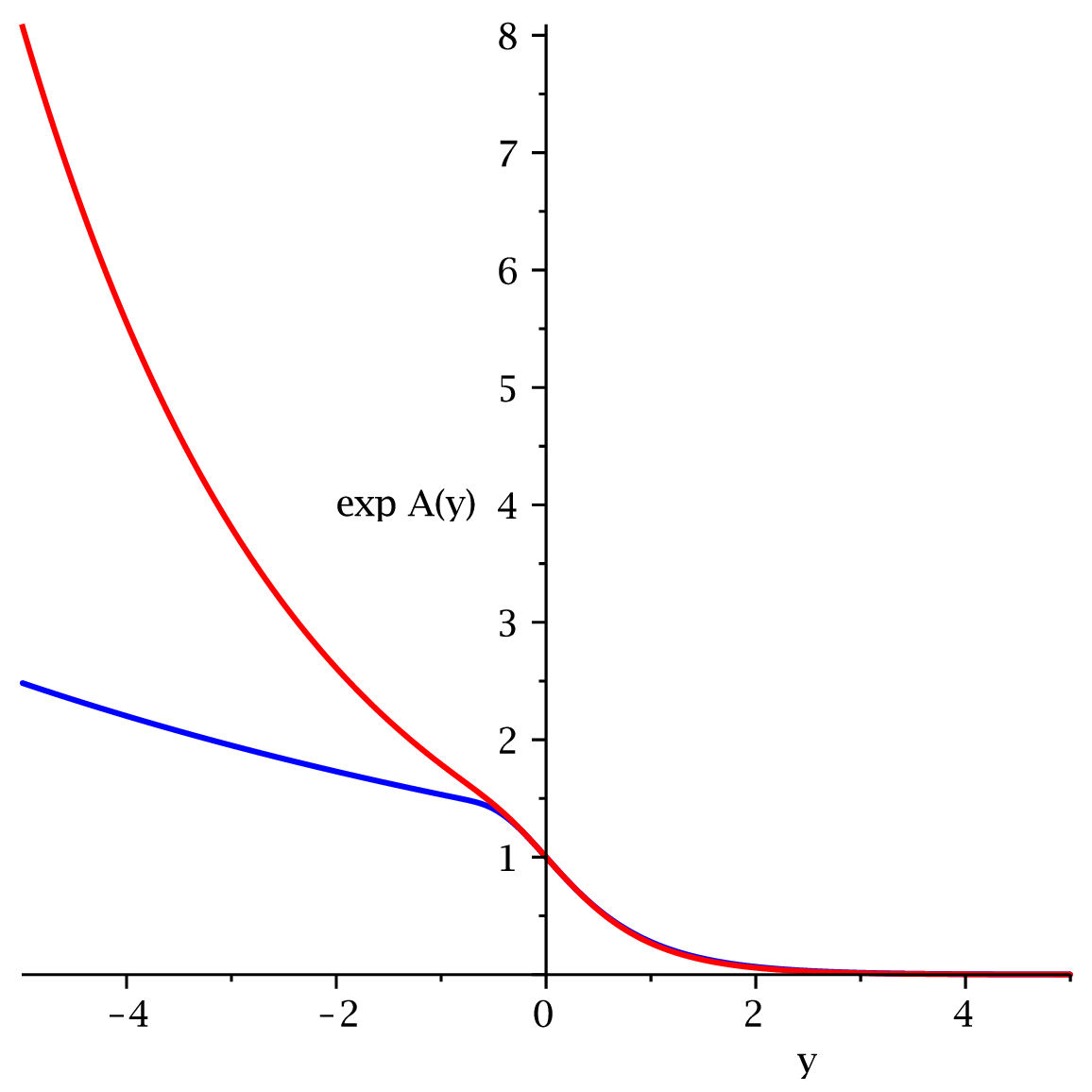}
\caption{The figure shows the warp factor $A(y)$ and scale factor $a(y)=\exp{A(y)}$ for the values $\beta=2.273$, with $\alpha=5/2$, $\eta=0.66$ (blue) and $\beta=3.125$, with $\alpha=5/2$, $\eta=0.48$ (red). The geometry is approaching those that connect AdS - AdS spacetimes (both curves). }
\label{p0-2}
\end{center}
\end{figure}

Let us now, consider the vacuum structure of the potentials, superpotential and the domain wall solutions. The potentials here considered develop structure of false vacua connected by domain walls.

From the discussions put forward in the last section, one might wonder whether these vacua decay in the context of the Horndeski gravity. We can invoke the Coleman-de Luccia (CdL) condition that allows vacuum decay. The well-know CdL condition relates the vacuum potentials to the domain wall surface tension $\sigma$ as follows \cite{Coleman:1980aw}
\begin{eqnarray}\label{CdL}
\varepsilon\geq\frac32\sigma^2,
\end{eqnarray}
where $\varepsilon\equiv\Delta \tilde{V}=\tilde{V}(\phi_+)-\tilde{V}(\phi_-)$. The violation of this inequality means that there is no vacuum decay.  Since we are dealing with AdS domain walls we can also write $\varepsilon$ in terms of the AdS lengths $\ell_+$ and $\ell_-$ --- see below.

Now, evaluating the potential \eqref{8} at the `supersymmetric' vacua $\phi_{vac}=\phi_\pm$ at which $\tilde{W}_\phi=0$, Fig.~\ref{W-Wp}, the scalar potential achieves the vacua of AdS spacetimes, that can be given in terms of the cosmological constant, that is
\begin{eqnarray}\label{CdL-2}
\tilde{V}(\phi_{vac})\equiv \Lambda_{AdS} =-\frac32 \tilde{W}^2(\phi_{vac}).
\end{eqnarray}
We are now able to explore the consequences of this result by substituting into the CdL condition \eqref{CdL} to obtain a condition on the domain wall tension that reads
\begin{eqnarray}\label{CdL-3}
\sigma^2&\leq&( \tilde{W}_- -\tilde{W}_+)^2
\end{eqnarray} 
or simply
\begin{eqnarray}\label{CdL-3.1}
\sigma\leq |\tilde{W}_- -\tilde{W}_+|,
\end{eqnarray} 
where we have assumed that $|\tilde{W}_-|>|\tilde{W}_+|$. On the other hand,  the domain wall surface density is given by \eqref{EF3-sigma} that in  its thin wall counterpart form \eqref{SE-8}, satisfies the inequality  
\begin{eqnarray}\label{CdL-4}
\sigma&\geq& |\tilde{W}_- -\tilde{W}_+|.
\end{eqnarray}
This Bogomol'nyi bound that relates the superpotential and the domain wall tension is  saturated within the use of the first order equations \eqref{fist}.
Notice that the Coleman-de Luccia bound \eqref{CdL-3.1} has opposite inequality sign as \eqref{CdL-4}. Thus, it is expected no vacuum decay in this context, which reveals that we have found infinite planar domain walls indeed. Furthermore, at the saturation limit of \eqref{CdL}, we conclude that all energy is converted to form the bubble wall, leaving no energy left to accelerate the wall to the speed of light,  asymptotically. Thus, the vacuum energy difference $\varepsilon$ approximates its minimum and the radius of the bubble wall associated with the false vacuum decay becomes very large. This precisely coincides with the configuration of static infinite planar domain walls separating vacua associated with different spacetimes \cite{Cvetic:1992st}.

\begin{figure}[!ht]
\begin{center}
\includegraphics[scale=0.35]{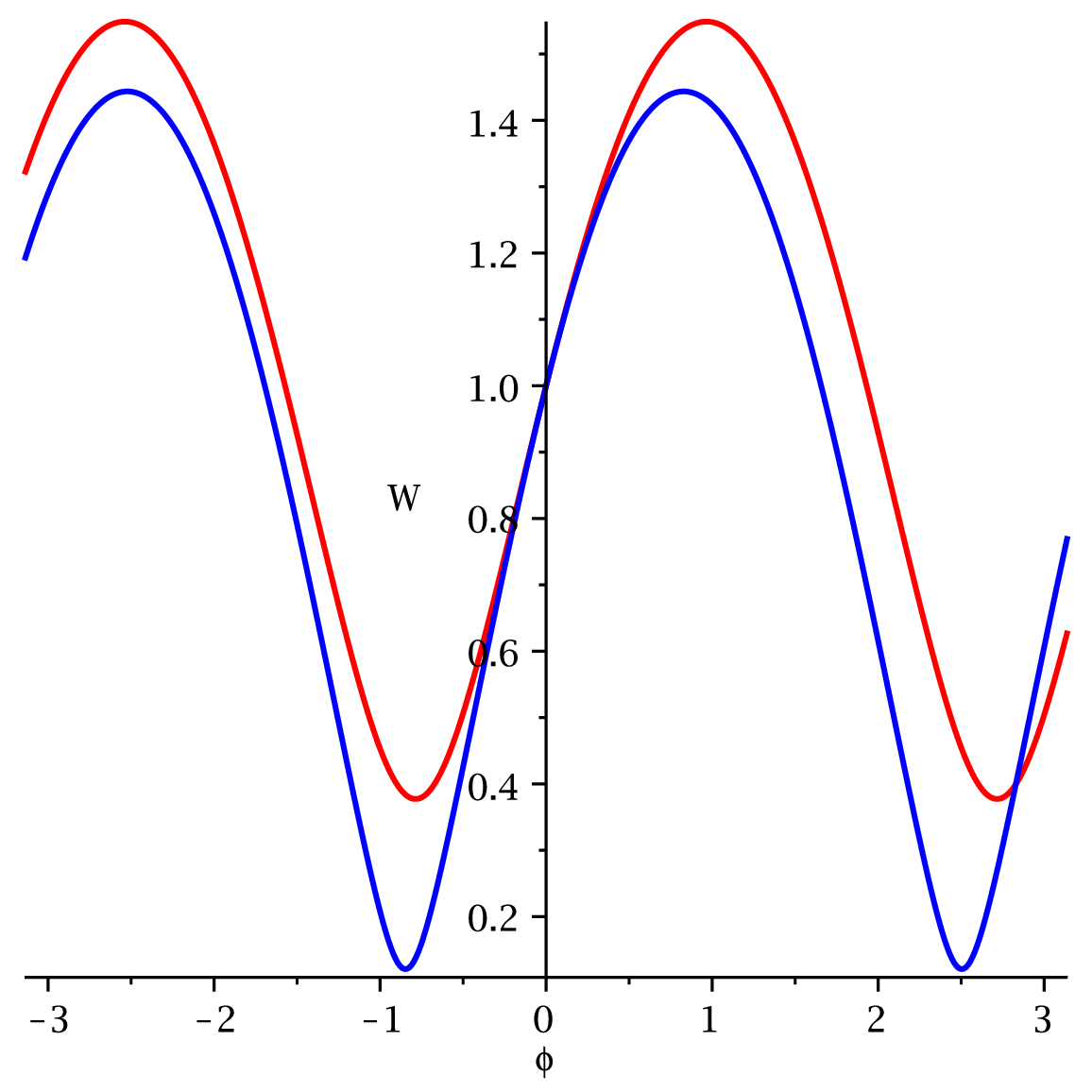}\qquad\qquad\qquad
\includegraphics[scale=0.35]{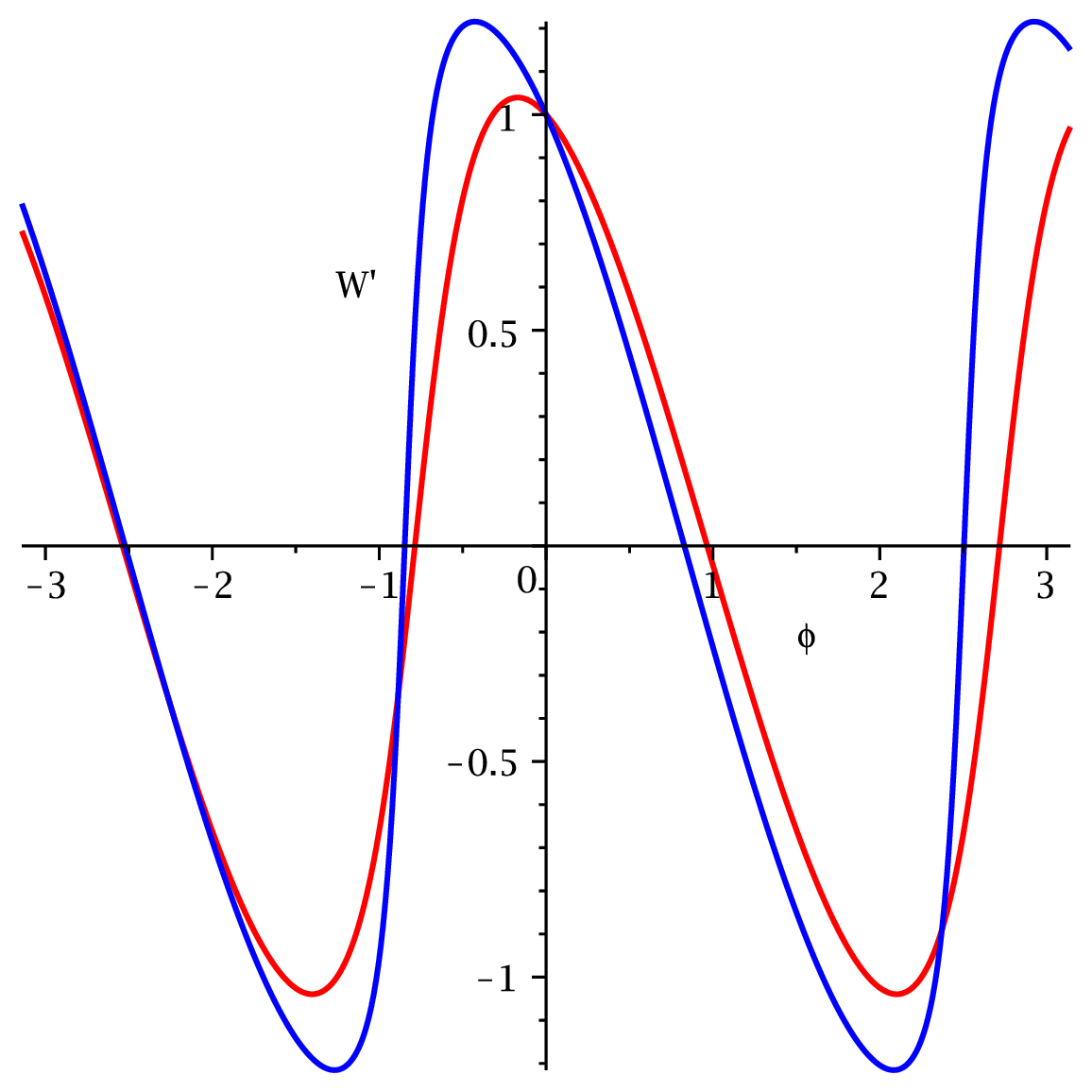}
\caption{The figure shows the superpotential and its derivative with respect to the scalar field for the values $\beta=2.273$, with $\alpha=5/2$, $\eta=0.66$ (blue) and $\beta=3.125$, with $\alpha=5/2$, $\eta=0.48$ (red). The `supersymmetric' vacua at $W'\equiv\tilde{W}_\phi=0$ are $\phi_{vac}=\phi_\pm \in (-1,1)$. See also Table \ref{Tb1}. They are interpolated by domain walls with the kink profiles given in Fig.~\ref{p0-1}. Notice that the superpotential enjoys the property $\tilde{W}(\phi)\neq 0$.}
\label{W-Wp}
\end{center}
\end{figure}
These computations are normally considered in the thin wall limit \cite{Coleman:1980aw} --- See also \cite{Ghosh:2021lua} for a recent comprehensive review. In the present study we can trust our computations because for the suitable choice of the parameters $\alpha$ and $\eta$ we can obtain numerical solutions that are in good agreement with the thin wall limit. For an important check of this behavior we can compare the domain wall thickness $\Delta$ with AdS radii $\ell_\pm$ in both examples previously considered according to the following criteria\footnote{This should be understood as a regime where the thickness is small enough to allow us to separate the bulk and domain wall contributions safely, as well-discussed in the subsection \ref{thin-wall-limit-subs}.} 
\begin{eqnarray}\label{thin-wall-limit}
r=\frac{2\Delta}{\ell_- + \ell_+}\equiv\frac13(\phi_+ - \phi_-)^2\frac{W_+ W_-}{W_-^2 -W_+^2}\ll 1,
\end{eqnarray}
where $\ell_- =1/W_-$ and $\ell_+ =1/W_+$. The domain wall thickness is self-consistent with the lower bound at \eqref{CdL-4} according to the definition $\Delta = (1/3) (\phi_+ - \phi_-)^2/\sigma$, for a fixed $\sigma$ regardless whether the domain wall is thin or not --- see \cite{Brito:2001hd,Bazeia:2004yw} for a detailed analysis on this issue. Alternatively, $\Delta$ can also be easily estimated by checking the kink derivative width at half maximum. This is particularly clear in the next section --- See Fig.~\ref{p1} up to the pre-factor $\eta$. Finally, in the Table \ref{Tb1} we show that according to the ratio $r$ \eqref{thin-wall-limit} our solutions are fairly in good agreement with the thin wall limit.

\begin{table}[!ht]
\begin{center}
\begin{tabular}{|c|c|c|c|c|c|c|} \hline
$\beta$&$\phi_+$&$\tilde{W}_-\equiv \tilde{W}(\phi_+)$&$\phi_-$&$\tilde{W}_+\equiv \tilde{W}(\phi_-)$&$|\tilde{W}_- -\tilde{W}_+|$&$r$\\ \hline
$2.273$&$0.8275$&$1.4435$&$-0.8473$&$0.1207$&$1.3228$&$0.0787$\\ \hline
$3.125$&$0.9638$&$1.5487$&$-0.7862$&$0.3771$&$1.1716$&$0.2642$\\ \hline
\end{tabular}
\end{center}
\caption{The table shows the superpotential at the vacua $\phi_\pm$ and the lower bound of domain walls tension for the values of $\beta$ considered in the present analysis. The ratio $r$ is in accord with the thin wall limit.}
\label{Tb1}
\end{table}

Before going to the next section some comments are in order. In the aforementioned analysis we focus on domain walls connecting AdS local minimum to AdS local maxima. This seems unusual, but AdS vacua can be local maxima of the potential indeed, as long as the Breitenlohner-Freedman (BF) bound is satisfied. In the present study, the fact that AdS vacua can appear as maxima is due to the second term of the potential \eqref{8} that brings a mixing of $\tilde{W}_\phi$ and $\tilde{W}$ such that the concavity of the critical points does not behave as in the usual manner. 
However, the domain walls are connecting `supersymmetric' vacua in the same way as in usual supergravity potentials in four dimensions \cite{Cvetic:1996vr,Cvetic:1993xe,Cvetic:1992st,Skenderis:2006jq} in the sense that the vacua satisfy the condition $\tilde{W}_\phi=0$. In the present theory, the behavior around these critical points is different as we can see in the graphs of the potentials in Figs.~\ref{p0-1}.  In Figs.~\ref{alpha-eta} we show how the concavity of the potential, at the critical point $\phi_+$ (right panel) changes as one moves along the parallel lines crossing the regions where the concavity of the potential changes (left panel). In the present example the brown line crosses the blue curve (for $\beta=2.273$) at $\alpha=1.455$ and $\eta=0.2$. At this point we plot the potential (brown curve) developing an opposite concavity at the critical point $\phi_+$ in relation to the potential (blue curve) previously shown in Fig.~\ref{p0-1}. Since the `supersymmetric' domain walls depend only on the form of the superpotential obtained via Eq.~\eqref{10} whose only parameter involved is $\beta$ the solution does not change regardless whether the potential changes its concavity at $\phi_+$ or not. This is, however, true as long as we keep crossing the lines where $\beta$ is fixed: the blue and red lines at Fig.~\ref{alpha-eta}. They are curves whose slopes are $\beta$, i.e., at  $\beta=2.273$ (blue curve) and  $\beta=3.125$ (red curve). In the shaded regions where the concavity is negative the BF bound $m^2\geq m_{BF}^2=-d^2/4$ can be easily satisfied. For example, for the potential developing maxima in $\phi_+$, as discussed above, Fig.~\ref{alpha-eta} (blue curve), with $\alpha=1.68$ and $\eta=0.3$, we find $m^2\equiv V''(\phi_+)=-1.539$, which is large enough to satisfy the BF bound for $d=3$. Finally, one can show that there are no allowed regions at which the critical point $\phi_-$ (right panel) changes concavity.
\begin{figure}[!ht]
\begin{center}
\includegraphics[scale=0.35]{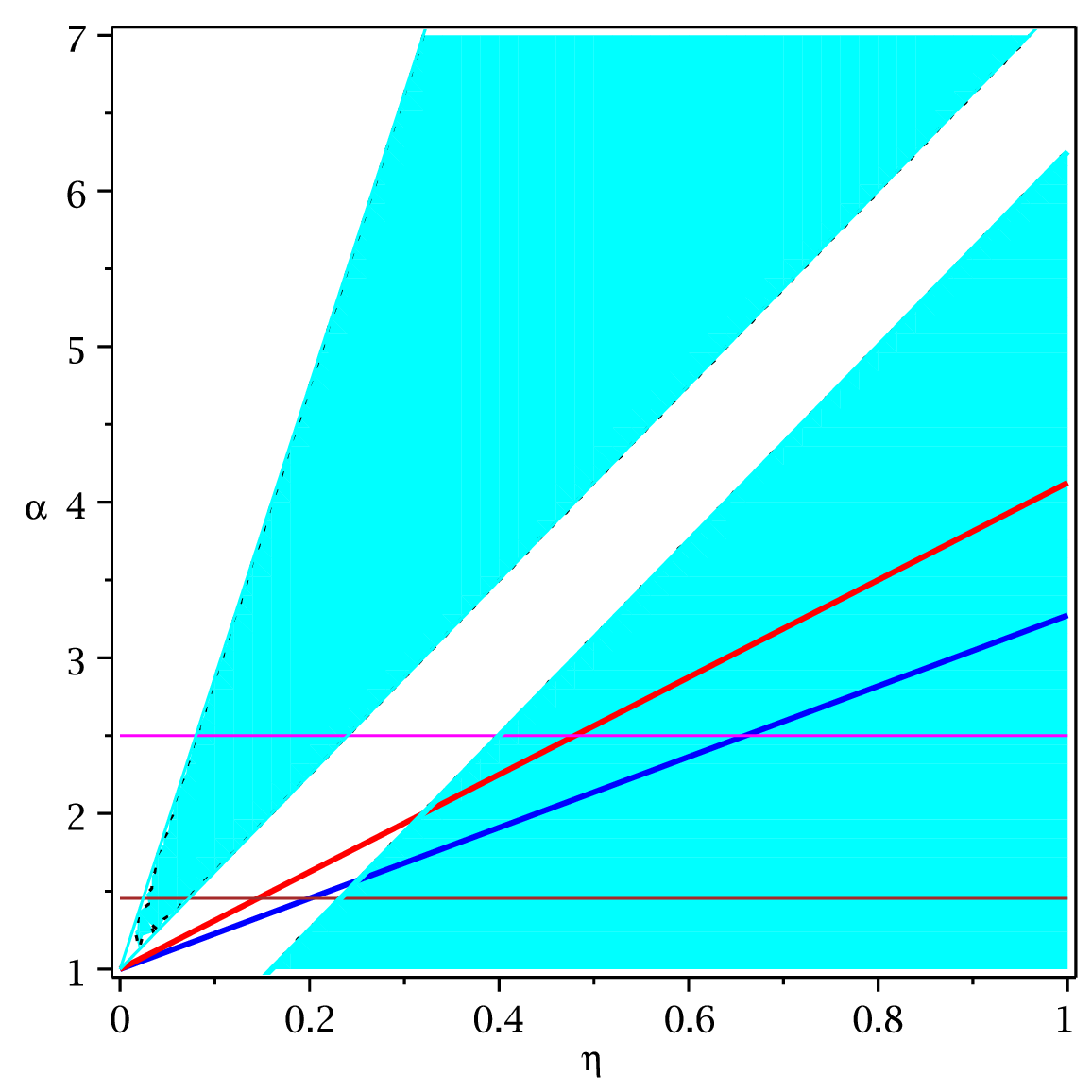}\qquad\qquad\qquad
\includegraphics[scale=0.35]{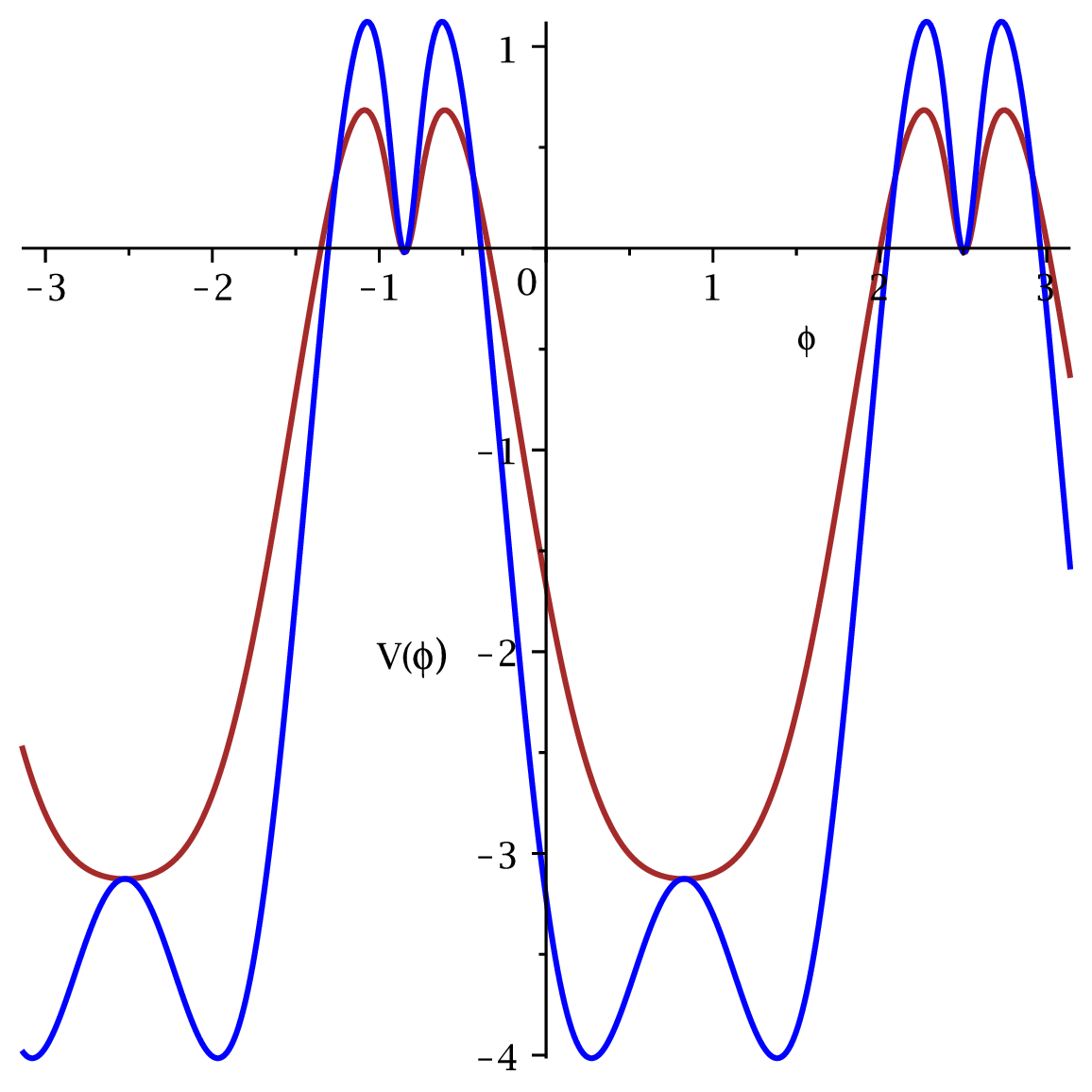}
\caption{The figure (left) shows the regions in the space of parameters where $V''(\phi_+)<0$ (the shaded regions) and $V''(\phi_+)>0$ (the white regions). Concerning the example with $\beta=2.273$  (blue curve) the horizontal lines cross the blue curve at $\alpha=5/2$ (potential with a maximum at $\phi_+$) and at $\alpha=1.455$   (potential with a minimum at $\phi_+$) (right).}
\label{alpha-eta}
\end{center}
\end{figure}

\subsection{Holographic RG flow}

It is also interesting to investigate the holographic renormalization group (RG) flow around the supersymmetric vacua.  In the RG flow we associate the scalar field $\phi$ as to a running coupling and the scale factor $a(y)=\exp{A(y)}$ as an scale of energy dictated by the $\tilde{\beta}(\phi)$-function, i.e., 
\begin{eqnarray}\label{RG-flow}
a\frac{d\phi(a)}{da}=\tilde{\beta}(\phi(a)).
\end{eqnarray}
By a quick check it is not difficult to rewrite Eq.~\eqref{RG-flow} in terms of Eqs.~\eqref{fist} as follows
\begin{eqnarray}\label{RG-flow2}
\tilde{\beta}(\phi)=-\frac{\tilde{W}'(\phi)}{\tilde{W}(\phi)}.
\end{eqnarray}
Now by expanding \eqref{RG-flow2} around the the supersymmetric vacua $\phi_*=\phi_\pm$ we have 
\begin{eqnarray}\label{RG-flow3}
\tilde{\beta}(\phi)=\tilde{\beta}(\phi_*)+\tilde{\beta}'(\phi_*) (\phi-\phi_*)+\cdot\cdot\cdot.
\end{eqnarray}
By recovering that the supersymmetric vacua satisfy $W'(\phi_*)=0$ and substituting \eqref{RG-flow3} at \eqref{RG-flow} we find the running coupling driven by the $\tilde{\beta}(\phi)$-function as 
\begin{eqnarray}\label{RG-flow4}
\phi(a)=\phi_* + c\, a^{\tilde{\beta}'(\phi_*)}.
\end{eqnarray}
Now we are able to identify the regimes of ultraviolet (UV) and infrared (IR) fixed points. For $\tilde{\beta}'(\phi_*)<0$ and $a\to\infty$, $\phi\to\phi_*$ we say that this vacuum is a UV stable fixed point, whereas for $\tilde{\beta}'(\phi_*)>0$, and $a\to0$, $\phi\to\phi_*$ we say that this vacuum is a IR stable fixed point. As can be seen in the Fig.~\ref{RG-flow-figs},  $\tilde{\beta}'(\phi_-)<0$ implies that the vacuum $\phi_-$ is a UV stable fixed point, while $\tilde{\beta}'(\phi_+)>0$ implies that the vacuum $\phi_+$ is a IR stable fixed point. Notice that the $\tilde{\beta}'(\phi)$-function in Fig.~\ref{RG-flow-figs} is normalized by $|\tilde{\beta}'(\phi_-)|$. 

Thus, in our examples shown in Fig.~\ref{p0-1} the `supersymmetric' domain wall solutions connect the false vacua (local minima) $\phi_-$ that correspond to UV stable fixed points to `true' vacua (local maxima/minima --- see Fig.~\ref{alpha-eta}) $\phi_+$  that correspond to IR stable fixed points. The latter case shows explicitly the relevance of choosing regions in the space of parameters to change the structure of vacua. And last but not least, we comment on the fact that for $\alpha=1$, the constraint equation \eqref{10} becomes a homogeneous equation and leads to
\begin{eqnarray}\label{constr}
W''(\phi_*)=-\frac32 W(\phi_*).
\end{eqnarray}
As a consequence the  $\tilde{\beta}'(\phi)$-function through the use of \eqref{RG-flow2} gives
\begin{eqnarray}
\tilde{\beta}'(\phi_*)=\frac32,
\end{eqnarray}
such that from \eqref{RG-flow4} we can see that this is a IR stable fixed point for $a\to0$. This kind of geometries with constraints similar to \eqref{constr} were well-addressed in \cite{Kallosh:2000tj} in the context of supergravity.

\begin{figure}[!ht]
\begin{center}
\includegraphics[scale=0.35]{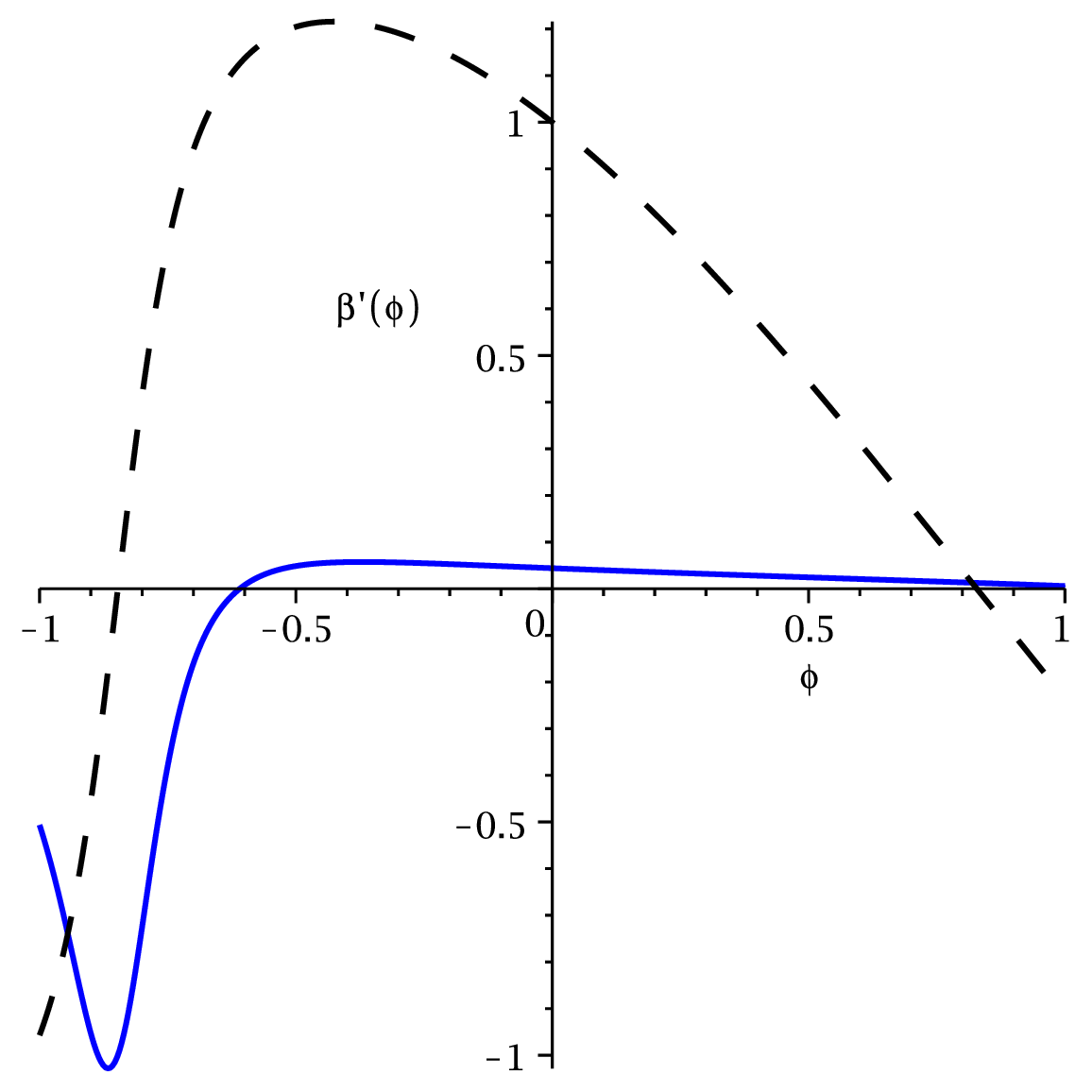}\qquad\qquad\qquad
\includegraphics[scale=0.35]{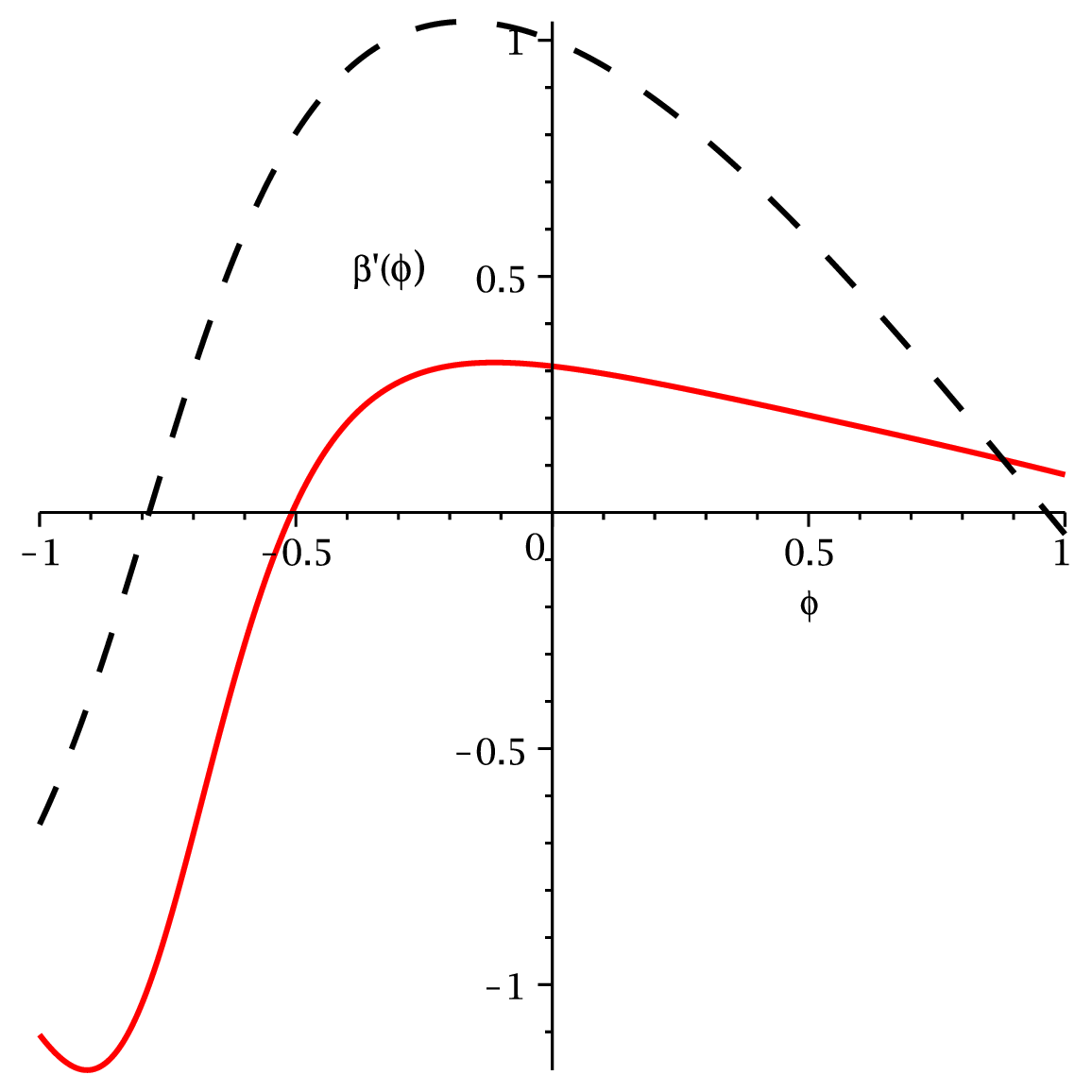}
\caption{The figure shows the normalized $\tilde{\beta}'(\phi_-)=-76.4020$ at $\phi_-=-0.8473$ (left) and $\tilde{\beta}'(\phi_-)=-9.483$ at $\phi_-=-0.7862$ (right) --- solid curves. The function changes the signal and achieves the vacua $\phi_+=0.8275$ and $\phi_+=0.9638$, becoming $\tilde{\beta}'(\phi_+)=0.9546$ and $\tilde{\beta}'(\phi_+)=0.8485$, respectively. The vacua are defined by $W'(\phi_\pm)=0$ --- dashed curves.}
\label{RG-flow-figs}
\end{center}
\end{figure}

\section{The domain wall/cosmology correspondence}\label{z3}
In this section, we evaluate the gravity fluctuations. By using the ADM formalism for Horndeski gravity presented in Ref.~\cite{Kob} for cosmological fluctuations and through the analytic continuation, we shall address the domain walls counterpart of linearized equations for tensor perturbations, shortly. In order to compute such equations, let us start considering the flat FLRW metric
\begin{equation}
ds^{2}=-dt^{2}+a^{2}(t)\delta_{ij}dx^{i}dx^{j},\label{5.1}
\end{equation}
where the scalar field depends on cosmic time only. It can be seen that \eqref{5.1} can be rewritten through the analytic continuation $W\to i\tilde{W}$, $H\to i\tilde{H}$, $t\to iy$ and $y\to it$ to achieve the domain wall metric (\ref{6}). The domain wall/cosmology correspondence also leads domain wall solutions to cosmological solutions \cite{Skenderis:2006jq}. The cosmological solutions are precisely kink-like solutions in terms of the time coordinate with the scalar potential suffering the sign changing, i.e., $V(\phi)\to - V(\phi)$ (this also asks the changing $\eta\to -\eta$ in \eqref{8}). In Figs.~\ref{p0-3} and \ref{p0-4} we show the analogous domain wall solutions and geometric  behavior in the cosmological scenario. We can see that now we have a geometry connecting de Sitter-de Sitter (dS-dS) spacetimes with diferente cosmological constants. As in the domain walls cases the vacua structure can be seen from the vacua at the scalar potentials in Fig.~\ref{p0-3} --- now the {\sl inflaton} field $\phi(t)$ rolls from the local minima on the right side to local maxima on the left side, after the potential suffering a phase transition, according to the mechanism revealed in Fig.~\ref{alpha-eta} (see discussion about the way the blue curve becomes the brown curve in the left panel.)

\begin{figure}[!ht]
\begin{center}
\includegraphics[scale=0.35]{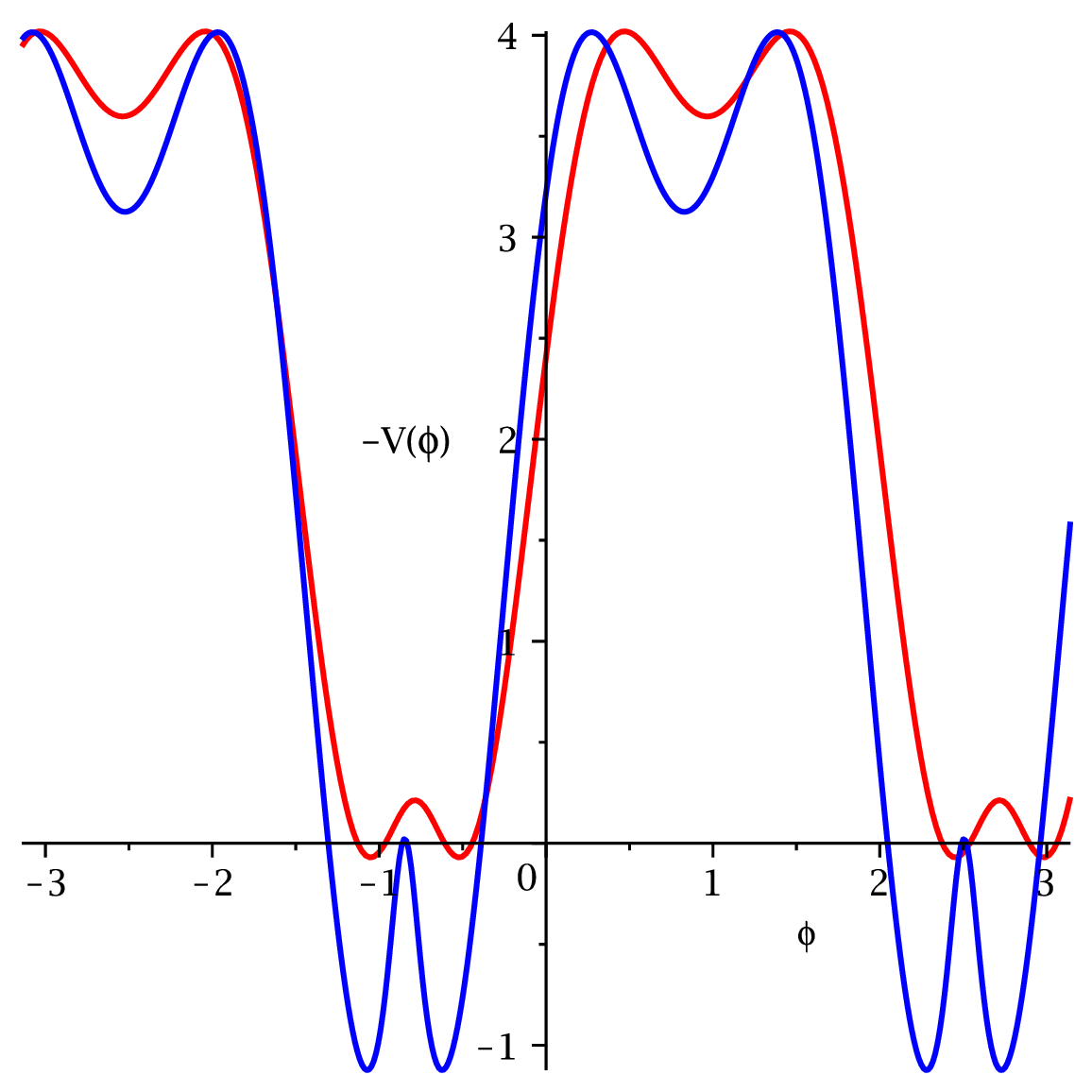}\qquad\qquad\qquad
\includegraphics[scale=0.35]{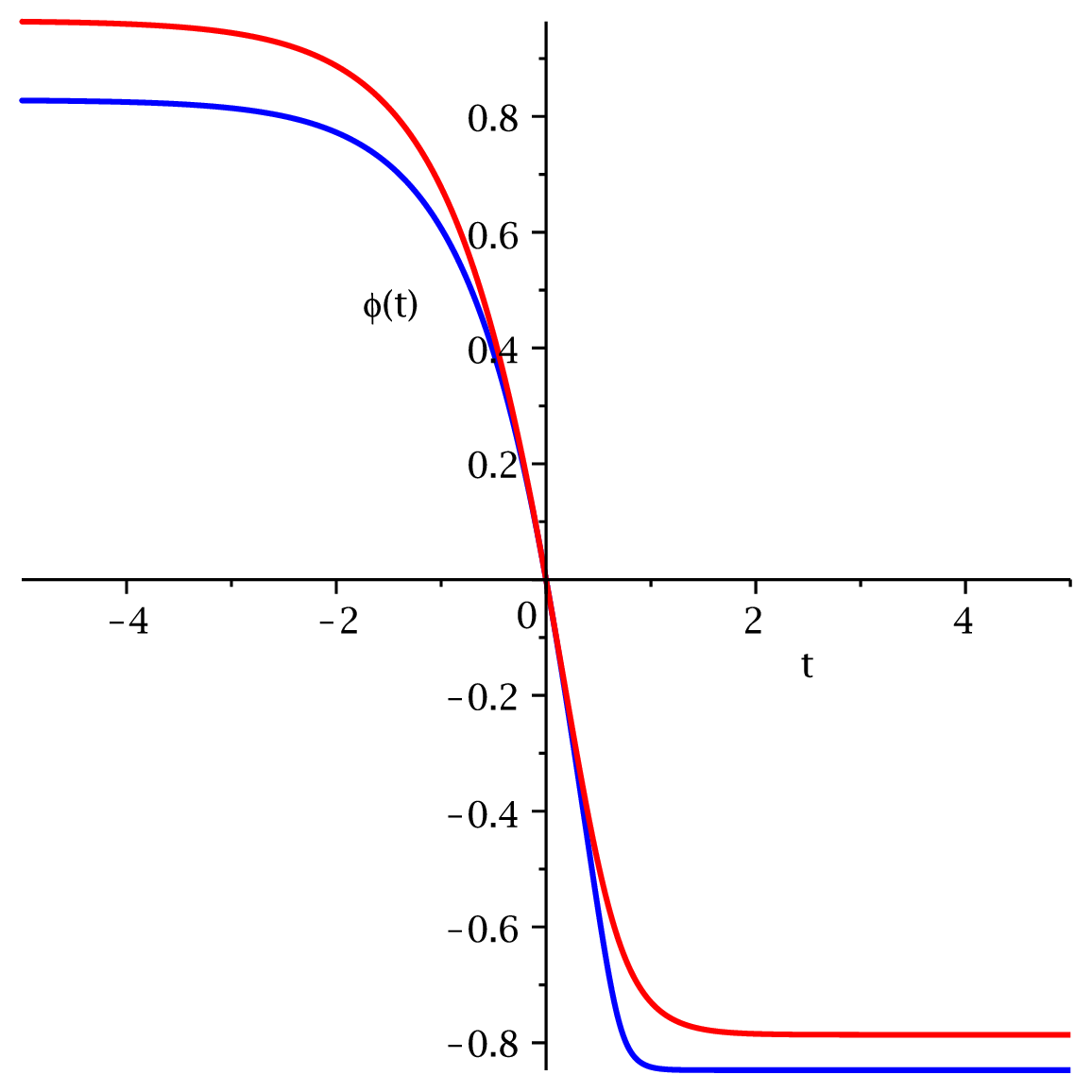}
\caption{The figure shows the scalar potential (\ref{8}) and kinks connecting the false vacua for the values $\beta=2.273$, with $\alpha=5/2$, $\eta=0.66$ (blue) and $\beta=3.125$, with $\alpha=5/2$, $\eta=0.48$ (red).}
\label{p0-3}
\end{center}
\end{figure}

\begin{figure}[!ht]
\begin{center}
\includegraphics[scale=0.35]{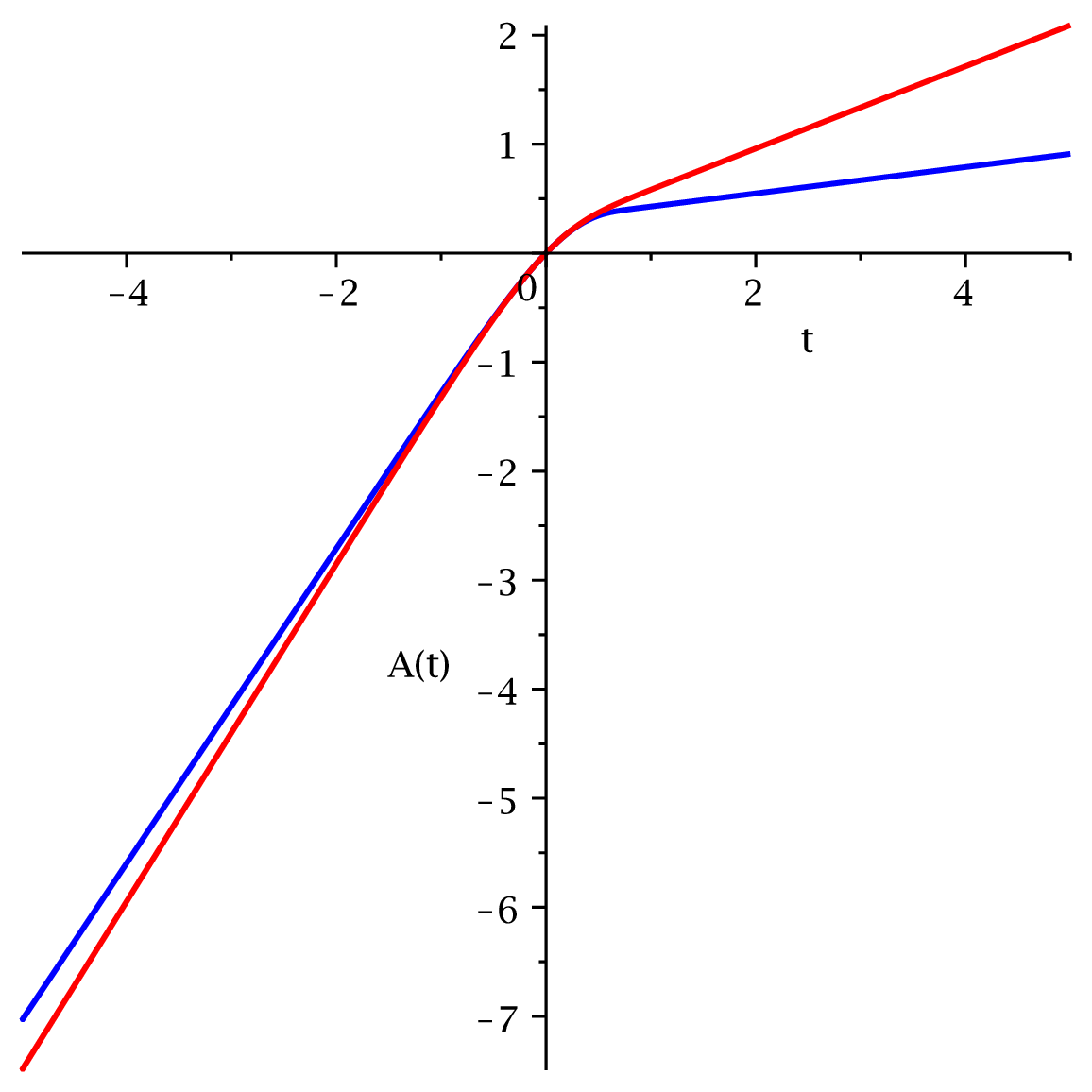}\qquad\qquad\qquad
\includegraphics[scale=0.35]{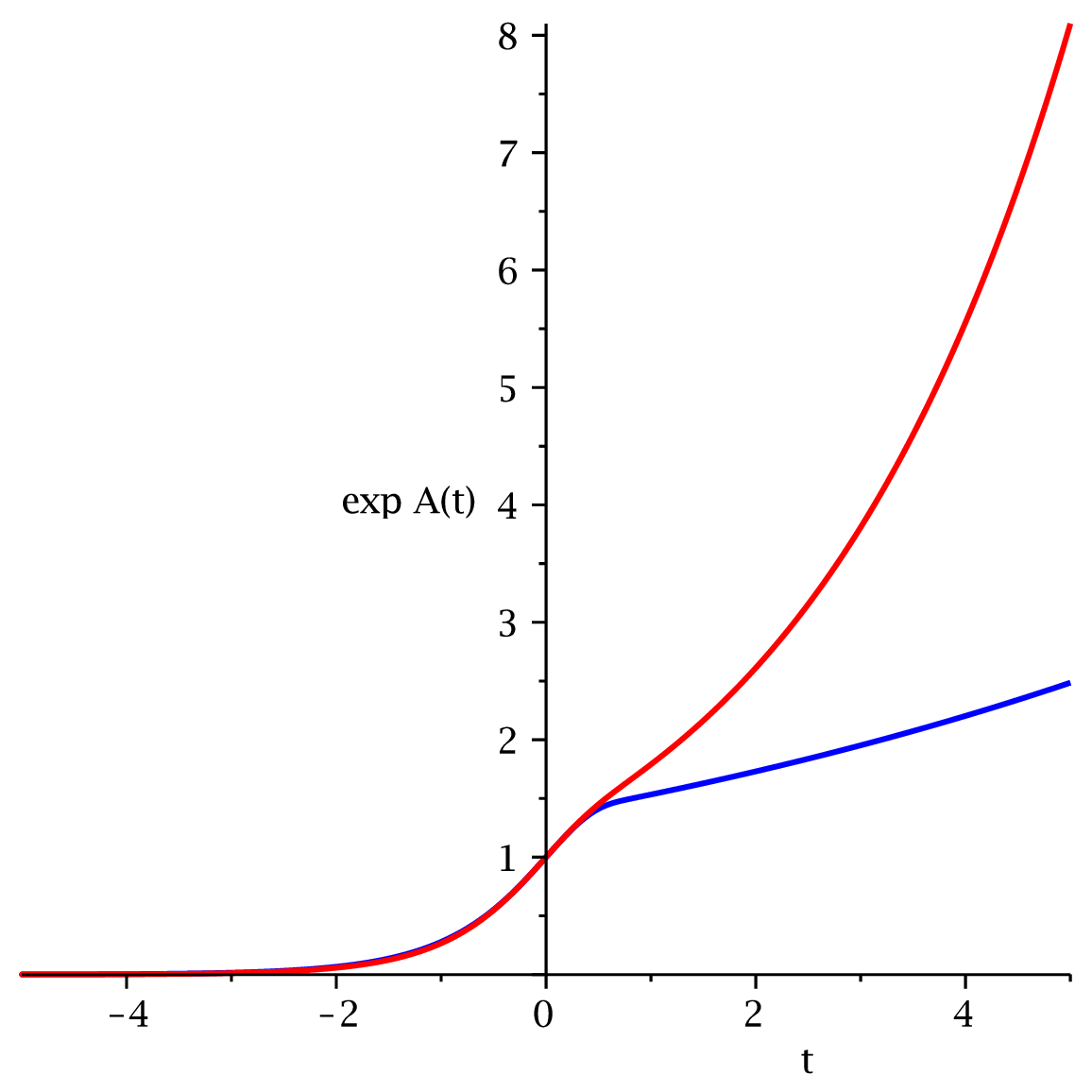}
\caption{The figure shows the warp factor $A(t)$ and scale factor $a(t)=\exp{A(t)}$ for the values $\beta=2.273$, with $\alpha=5/2$, $\eta=0.66$ (blue) and $\beta=3.125$, with $\alpha=5/2$, $\eta=0.48$ (red). The geometry is approaching those that connect dS - dS spacetimes (both curves). }
\label{p0-4}
\end{center}
\end{figure}

For the sake of completeness, and to address the issues concerning linear stability it is also interesting to establish the domain wall correspondence of cosmological tensor fluctuations. Since tensor fluctuations decouple from scalar perturbations we shall simply focus on the first type. Thus from the action for tensor perturbations, up to second order, in Horndeski gravity (\ref{01}) at ADM formalism
\begin{eqnarray}
S^{(2)}_{\rm tensor}=\frac18\int{dtd^3xa^3\left[G_T \dot{h}_{ij}^2-\frac{F_T}{a^2}(\partial_k h_{ij})^2\right]},
\end{eqnarray}
and TT (transverse and traceless) gauge, $\partial^i h_{ij}=0, h_i^i=0$, we find the equations for the tensor fluctuations given by
\begin{eqnarray}
T(t)\ddot h_{ij}+B(t)\dot{h}_{ij}-a^{-2}\nabla^2 h_{ij}=0,\label{T1}
\end{eqnarray}
where
\begin{eqnarray}
T(t)\equiv\frac{G_T}{F_T}, \:\:\: B(t)\equiv\frac{\dot{G}_T+3HG_T}{F_T}.
\end{eqnarray}
The explicit form of the coefficients $G_T$ and $F_T$ reads \cite{Kob}
\begin{eqnarray}
&& G_T\equiv 2\left[G_4-2XG_{4X}-X\left(H\dot\phi G_{5X}-G_{5\phi}\right)\right]=2\left(G_4+X G_{5\phi}\right),\\
&& F_T\equiv 2\left[G_4-X\left(\ddot\phi G_{5X}+G_{5\phi}\right)\right]=2\left(G_4-X G_{5\phi}\right),
\end{eqnarray}
where in the last step the coefficients simplify under our choice of $G_i$'s in the action (\ref{1}). Thus we find
\begin{eqnarray}
T(t)=\frac{4\kappa-\eta\dot{\phi}^2}{4\kappa+\eta\dot{\phi}^2}, \:\:\: B(t)=\frac{3H(4\kappa-\eta\dot{\phi}^2)-2\eta\dot{\phi}\ddot{\phi}}{4\kappa+\eta\dot{\phi}^2}.
\end{eqnarray}
It is also interesting to notice that the gravitational wave speed is defined in terms of $1/T(t)$, that is
\begin{eqnarray}
c_{\rm GW}^2\equiv\frac{F_T}{G_T}=\left.\frac{4\kappa+\eta\dot{\phi}^2}{4\kappa-\eta\dot{\phi}^2}\right|_{t\to t_0}\to1=c_{\rm light}^2,\label{light}
\end{eqnarray}
behaves well in accord with the measurement of the events gravitational waves GW170817 and the $\gamma$-ray burst GRB170817A detected almost simultaneously, because of the suitable evolution of the scalar field in large times --- as depicted in Fig.~\ref{p0-3}, $t_0=4$. Large times correspond to low redshifts which correspond to the energy scale of the dark energy and GW170817/GRB170817A events. {Furthermore, as shown in Fig.~\ref{p1}, the stability is ensured because since $\eta\dot{\phi}^{2}(t)<1$, for $\beta=2.273$, with $\alpha=5/2$, $\eta=0.66$ (blue curve) and $\beta=3.125$, with $\alpha=5/2$, $\eta=0.48$ (red curve) then $4\kappa>\eta\dot{\phi}^2$, which implies that $c_{\rm GW}^2$ in Eq.~\eqref{light} is a positive number.  These analyses were performed through numerical methods with the following conditions: $\tilde{W}(0)=1$, $\dot{\tilde{W}}(0)=1$ and $\phi(0)=0$.  We can see that even with the restriction $c_{GW}=c_{light}$ as presented by \cite{Charmousis:2011bf,Charmousis:2011ea,Babichev:2017lmw,Starobinsky:2016kua,Bruneton:2012zk}, these models still persists as important alternative cosmological scenarios. In this sense, we have that the energy-momentum tensor of the scalar field almost perfectly counterbalances the large bare cosmological constant that is supposed to be present in the Lagrangian. In this way, the observable accelerated expansion of the Universe becomes consistent with a small effective cosmological constant.}

\begin{figure}[!ht]
\begin{center}
\includegraphics[scale=0.36]{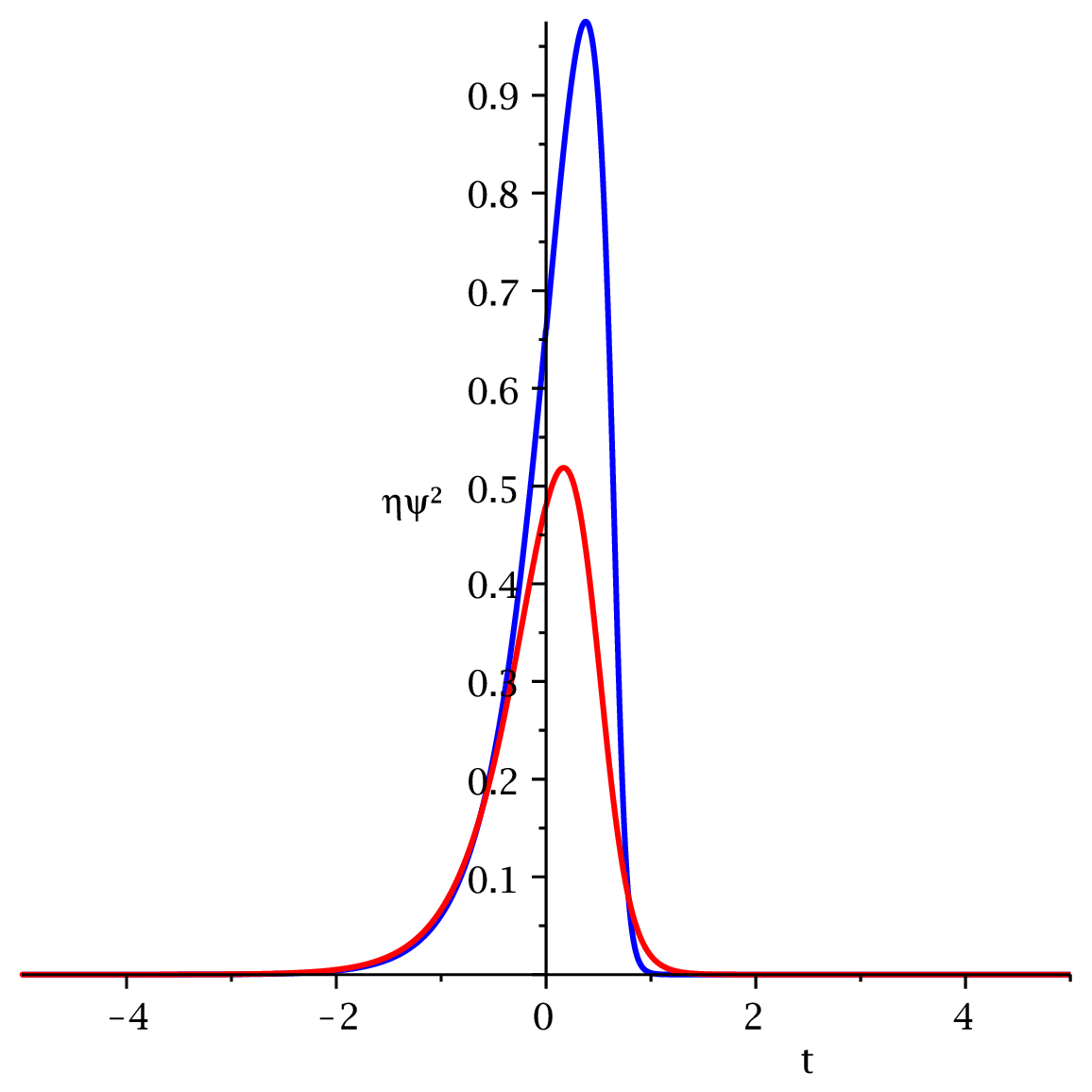}
\caption{The figure shows $\eta\psi^2\equiv\eta\dot{\phi}^{2}(t)$ for the values $\beta=2.273$, with $\alpha=5/2$, $\eta=0.66$ (blue) and $\beta=3.125$, with $\alpha=5/2$, $\eta=0.48$ (red). The maximum achieved by the blue curve is 0.975 at $t=0.378$.}
\label{p1}
\end{center}
\end{figure}

Now making use of the aforementioned analytic continuation one can show that the tensor perturbations and considering the $h_{\mu\nu}$ transverse traceless (TT) tensor perturbation with $\eta^{^{\mu\lambda}}\partial_{\lambda}h_{\mu\nu}=0$, $h\equiv\eta^{\mu\nu}h_{\mu\nu}=0$ in the metric (\ref{5.1}) for domain wall solutions are given by
\begin{eqnarray}
\tilde{T}(y)h_{\mu\nu}''+\tilde{B}(y){h}_{\mu\nu}'+a(y)^{-2}\square h_{\mu\nu}=0.\label{T2}
\end{eqnarray}
Our coefficients are now given in terms of the spatial coordinate due to the analytic continuation and are written as
\begin{eqnarray}
\tilde{T}(y)=\frac{4\kappa+\eta{\phi'}^2}{4\kappa-\eta{\phi'}^2}, \:\:\: \tilde{B}(y)=\frac{3\tilde{H}(4\kappa+\eta {\phi'}^2)+2\eta\phi'\phi''}{4\kappa-\eta{\phi'}^2}.
\end{eqnarray}
In fact, the equations (\ref{T1}) and (\ref{T2}) are formally the same as we can see by applying the analytic continuation $t\to iy$ and $y\to it$. This formalism for gravitational fluctuations around domain walls solutions in four dimensions is the analogue of the Randall-Sundrum scenario in five-dimensions \cite{Randall:1999vf,Randall:1999ee}.  We can use these fluctuations to address similar phenomenon of gravity localization in lower dimensional objects such as 2-branes embedded in higher dimensional structures as in cascading gravity addressed recently \cite{Hao:2014tsa,Bazeia:2014xfa} that we shall reconsider in future investigations in the Horndeski gravity context.

\section{Conclusions}\label{z4}

We have investigated BPS domain walls in Horndeski gravity. The superpotential is constrained by a differential equation in the space of the scalar field. The domain wall solutions found in the present study are connected by non-degenerate vacua. The vacua can be in general Minkowski ou AdS type vacua, depending on the choice of parameters. In the present analysis we decide to focus on AdS vacua only and have discussed the general possibility of vacuum decay. We have shown,  however, that at thin wall limit, the false vacuum cannot decay because the Coleman-de Luccia bound \cite{Coleman:1980aw} is not satisfied. As a consequence the bubble becomes very large and coincides with flat and static BPS domain walls. This is precisely what is found in supergravity theories as shown long ago \cite{Cvetic:1996vr,Cvetic:1993xe,Cvetic:1992st}. Now we have found the same effect in Horndeski gravity under certain limit of parameters of the theory. Under analytic continuation the study can also be extended to de Sitter vacua.

We also investigate the holographic renormalization group flow around the supersymmetric vacua. We notice that although the system develops vacua structure similar to double sine-Gordon (SG) systems, differently our system can develop domain walls that connects minimum to maximum vacua, i.e. it can connect UV to IR stable fixed points without skip intermediate vacua as it happens normally in double SG models. This can be understood as a consequence of the particle developing high velocity since the slope from one vacuum to another is very steep. As a consequence the gravitational friction terms act more strongly such that the particle cannot save energy enough to skip intermediate maxima and achieve the next UV stable fixed point. This is a point to be further addressed in the limit where the supersymmetric theory is slightly perturbed by $\eta$-term that normally appears in the full theory without taking the thin wall limit. We are considering these issues and shall put forward elsewhere shortly.

\acknowledgments

We would like to thank CNPq and CAPES for partial financial support. FAB acknowledges support from CNPq (Grant No. 309092/2022-19) and PRONEX/CNPq/FAPESQ-PB (Grant No. 165/2018), for partial financial support.  We also acknowledge the anonymous referee for important comments that helped us to improve the paper.

\end{document}